%% file: paper-sigplan.tex
\documentclass[conference,9pt]{IEEEtran}
\pdfoutput=1
\usepackage[cmex10]{amsmath}
\usepackage{amsthm,amssymb,latexsym}
\usepackage{upgreek}
\usepackage{graphicx}
\usepackage{bmpsize}
\usepackage{wasysym}
\usepackage{code}
\usepackage{comment}
\usepackage{multirow}
\usepackage[all]{xy}
\usepackage{url}
\usepackage{textcomp}
\usepackage{wrapfig}
\usepackage{lipsum}
\usepackage{stmaryrd}
\usepackage{mathtools}
\usepackage{subcaption}
\usepackage{relsize}
\usepackage{fancyvrb}
\usepackage[linesnumbered,ruled,vlined]{algorithm2e}
\usepackage{paralist}

\input{matt-might-macros}

\input{local-macros}


\mathchardef\mhyphen="2D

\newcommand{\subparagraph}{}
\renewcommand{\paragraph}[1]{\textbf{#1}:~}

\begin{document}
        
%% \special{papersize=8.5in,11in}
%% \setlength{\pdfpageheight}{\paperheight}
%% \setlength{\pdfpagewidth}{\paperwidth}

%% \titlebanner{\mytitlebanner}        % These are ignored unless
%% \preprintfooter{\footertitle}       % 'preprint' option specified.

\title{Pruning, Pushdown Exception-Flow Analysis}
%% \subtitle{\mysubtitle}

\author{
  \IEEEauthorblockN{Shuying Liang}
  \IEEEauthorblockA{University of Utah\\
    liangsy@cs.utah.edu}
  \and
  \IEEEauthorblockN{Weibin Sun}
  \IEEEauthorblockA{University of Utah\\
    wbsun@cs.utah.edu}
  \and
  \IEEEauthorblockN{Matthew Might}
  \IEEEauthorblockA{University of Utah\\
    might@cs.utah.edu}
  \and
  \IEEEauthorblockN{Andy Keep}
  \IEEEauthorblockA{University of Utah\\
    andy.keep@gmail.com}
  \and
  \IEEEauthorblockN{David Van Horn}
  \IEEEauthorblockA{University of Maryland\\
    dvanhorn@cs.umd.edu}}

\maketitle

\setlength{\abovedisplayskip}{2pt}
\setlength{\abovedisplayshortskip}{1pt}
\setlength{\belowdisplayskip}{2pt}
\setlength{\belowdisplayshortskip}{1pt}

\input{abstract2}

\input{intro2}

\input{content2}

{\small
\bibliographystyle{abbrv}
\bibliography{local,shuyingliang}
}

\end{document}

%% file: matt-might-macros.tex
\newcommand{\ie}{\emph{i.e.}}
\newcommand{\eg}{\emph{e.g.}}
\newcommand{\etc}{\emph{etc.}}
\newcommand{\etal}{\emph{et al.}}

\newcommand{\syn}[1]{\mathsf{#1}}
\newcommand{\var}[1]{\mathit{#1}}
\newcommand{\s}[1]{\mathit{#1}}

\newcommand{\parto}{\rightharpoonup}

\newcommand{\setbuild}[2]{\left\{ #1 : #2\right\}}

\newcommand{\Pow}[1]{{\mathcal{P}\left(#1\right)}}
\newcommand{\PowSm}[1]{{\mathcal{P}(#1)}}

\newcommand{\union}{\cup}

\newcommand{\vect}[1]{\langle #1\rangle}

\newcommand{\To}{\mathrel{\Rightarrow}}

\newcommand{\join}{\sqcup}

\newcommand{\sembr}[1]{\ensuremath{[\![{#1}]\!]}}

\newcommand{\opor}{\mathrel{|}}

\newcommand{\produces}{\mathrel{::=}}

\newcommand{\vv}{v}

\newcommand{\ttfs}{\mbox{\tt .}}

\newcommand{\ttlp}{\mbox{\tt (}}
\newcommand{\ttrp}{\mbox{\tt )}}
\newcommand{\ttlc}{\mbox{\tt \{}}
\newcommand{\ttrc}{\mbox{\tt \}}}

\newcommand{\ttsc}{\mbox{\tt ;}}

\newcommand{\stmt}{s}

\newcommand{\lab}{\ell}

\newcommand{\expr}{e}

\newcommand{\op}{{\var{op}}}

\newcommand{\State}{\Sigma}
\newcommand{\state}{\varsigma}

\newcommand{\Den}{D} % Denotable values
\newcommand{\den}{d}
\newcommand{\obj}{o}

\newcommand{\store}{\sigma}

\newcommand{\cont}{\kappa}

\newcommand{\alloc}{\mathit{alloc}}

\newcommand{\addr}{a}

\newcommand{\val}{\var{val}}

\newcommand{\tm}{t}
\newcommand{\tick}{{{tick}}}

\newcommand{\Lives}{\mathit{Lives}}

\newcommand{\aTo}{\leadsto}

\newcommand{\sa}[1]{\widehat{\mathit{#1}}}

\newcommand{\aqstate}{{\hat{q}}}

\newcommand{\aState}{{\hat{\Sigma}}}
\newcommand{\astate}{{\hat{\varsigma}}}

\newcommand{\aDen}{\hat{D}} % Denotable values

\newcommand{\astore}{{\hat{\sigma}}}

\newcommand{\aobj}{{\hat \obj}}

\newcommand{\acont}{{\hat{\kappa}}}

\newcommand{\aden}{{\hat{d}}}

\newcommand{\aaddr}{{\hat{\addr}}}

\newcommand{\aval}{{\widehat{\var{val}}}}

\newcommand{\atick}{{\widehat{tick}}}
\newcommand{\atm}{{\hat t}}

%% file: local-macros.tex
\newcommand{\methodDef}{M}
\newcommand{\constDef}{K}

\newcommand{\classform}[3]{
  \mathtt{class}\; 
  #1\;
  \mathtt{extends}\;
  #2\;
  \ttlc
  #3
  \ttrc
}
\newcommand{\con}

\newcommand{\className}{C}
\newcommand{\fieldName}{f}
\newcommand{\methodName}{m}

\newcommand{\MethodLookup}{{\mathcal{M}}}
\newcommand{\FetchRuctor}{{\mathcal{C}}}
\newcommand{\Ructor}{{\mathcal{K}}}

\newcommand{\ssucc}{\mathit{succ}}

\newcommand{\phrame}{\phi}
\newcommand{\aphrame}{\hat{\phi}}

\newcommand{\callframe}{\chi}
\newcommand{\acallframe}{\widehat{\chi}}
\newcommand{\exnframe}{\eta}
\newcommand{\aexnframe}{\widehat{\eta}}

\newcommand{\aCollect}{{\hat{G}}}

\newcommand{\aCallFrame}{\widehat{\mathit{CallFrame}}}

\newcommand{\fp}{\s{fp}}
\newcommand{\afp}{\hat{\mathit{fp}}}
\newcommand{\objp}{\s{op}}
\newcommand{\aobjp}{\sa{op}}

\newcommand{\ObjectPointer}{\s{ObjectPointer}}
\newcommand{\aObjectPointer}{\widehat{\ObjectPointer}}

\newcommand{\FramePointer}{\s{FramePointer}}
\newcommand{\aFramePointer}{\widehat{\FramePointer}}

\DeclareMathOperator*{\areaches}{\rightarrowtriangle}
\newcommand{\DSG}{\mathit{DSG}}
\newcommand{\IECG}{\mathit{IECG}}
\newcommand{\EpsPreds}{\overleftarrow{G}}
\newcommand{\EpsSuccs}{\overrightarrow{G}}

\newcommand{\TopFrames}{\overrightarrow{\mathit{TF}}}
\newcommand{\PossibleStackFrames}{\overrightarrow{\mathit{PSF}}}

\newcommand{\PredsForPush}{\overleftarrow{\mathit{PFP}}}
\newcommand{\NonEpsPreds}{\overleftarrow{\mathit{NEP}}}
\newcommand{\ToVisit}{{W}}
\newcommand{\deltaEdges}{\varDelta{E}}
\newcommand{\deltaStates}{\varDelta{S}}

\newcommand{\akont}{\hat{\kappa}}
\newcommand{\wk}{\sqsubseteq}

%% file: abstract2.tex
\begin{abstract}

Statically reasoning in the presence of exceptions and about the
effects of exceptions is challenging: exception-flows are mutually
determined by traditional control-flow and points-to
analyses. We tackle the challenge of analyzing exception-flows from
two angles. First, from the angle of pruning control-flows (both
normal and exceptional), we derive a pushdown framework for an
object-oriented language with full-featured exceptions. Unlike
traditional analyses, it allows precise matching of throwers to catchers.
Second, from the angle of pruning points-to information, we generalize
\textit{abstract} garbage collection to object-oriented programs and enhance it
with liveness analysis. We then seamlessly weave the techniques into
enhanced reachability computation, yielding highly precise
exception-flow analysis, without becoming intractable, even for large
applications. We evaluate our pruned, pushdown exception-flow
analysis, comparing it with an established analysis on large scale
standard Java benchmarks. The results show that our analysis
\textit{significantly} improves analysis precision over traditional analysis
within a reasonable analysis time.

\end{abstract}

%% file: intro2.tex
\section{Introduction}\label{sec:intro}
Exceptions are not exceptional enough.
They pervade the control-flow
structure of modern object-oriented programs.
An exception indicates an error occurred during program execution.
Exceptions are resolved by locating code specified by the programmer for
handling the exception (an exception handler) and executing this code.

\begin{figure}
 \centering
 \includegraphics[width=0.35\textwidth]{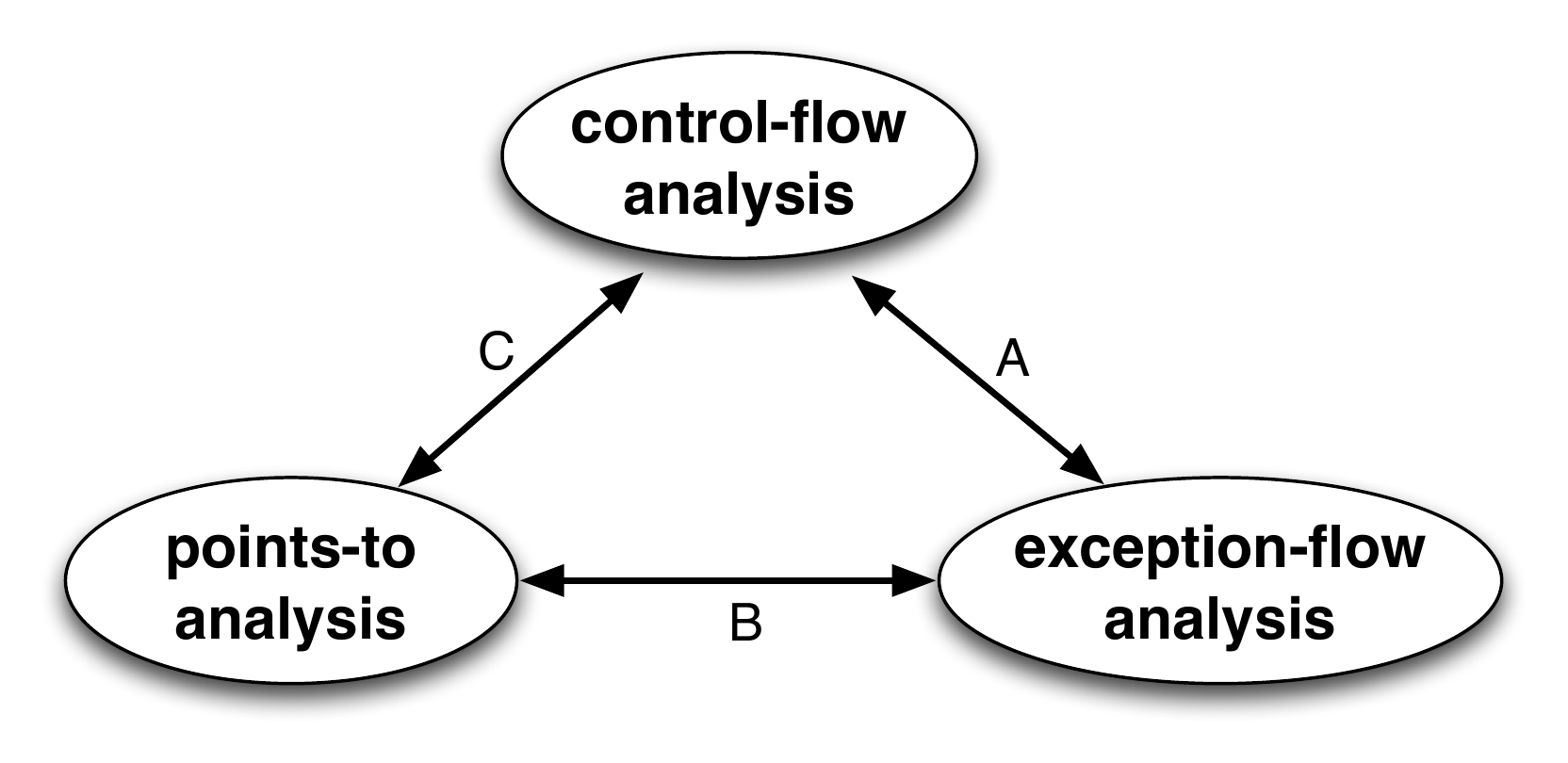}
 \caption{\label{fig:tri} Relationship among exception-flow analysis, control-flow analysis and points-to analysis.}
 \end{figure}

 This language feature is designed to ensure software robustness and reliability.
 Ironically, Android malware is exploiting it to leak  private sensitive information to the Internet through exception handlers~\cite{shuyingliang:Liang:2013:pushdownandroid}.
Analyzing the behavior of programs in the presence of exceptions is important to detect such vulnerabilities. 
However,  exception-flow analysis is challenging, because it
depends upon control-flow analysis and points-to analysis, which are
themselves mutually dependent, as illustrated in Figure~\ref{fig:tri}.

In Figure~\ref{fig:tri}, edge A denotes the mutual dependence between exception-flow analysis and traditional control-flow analysis (CFA).
CFA traditionally analyzes which methods can be invoked at each call-site.
Exception-flow analysis refers to the control-flow that is introduced when throwing exceptions~\cite{Bravenboer:2009:Exceptions}.
Intuitively, throwing an exception behaves like a global \texttt{goto} statement,
 in that it introduces additional, complex, inter-procedural control flow into the program. 
This  makes it difficult to reason about feasible run-time paths using 
 traditional CFA.
Similarly,  infeasible call and return flows can cause 
spurious paths between throw statements and catch blocks.
The following simple example demonstrates this:
%% \footnotesize
\begin{center}
	%% \begin{small}
	\begin{Verbatim}[fontsize=\relsize{-1}]
try {
   maybeThrow();  // Call 1
} catch (Exception e) {  // Handler 1
   System.err.println("Got an exception"); 
}
maybeThrow();  // Call 2
\end{Verbatim}
%% \end{small}
\end{center}
%% \normalsize
 Under a monovariant abstraction like 0-CFA~\cite{shuyingliang:Shivers:1991:CFA}, 
 where the distinction between different invocations of the same procedure are lost, 
 it will seem as though exceptions
 thrown from {\tt Call 2} can be caught by {\tt Handler 1}.
 
Edge B in Figure~\ref{fig:tri} 
denotes the relationship between exception-flow analysis and points-to analysis.  
Points-to analysis 
computes which abstract objects (with respect to allocation sites, calling contexts, etc.) 
a program variable or register can point to.
Points-to analysis affects exception-flow analysis, because the type of the
exception at a throw site determines which catch block will be executed.
That is to say, exception-flow analysis requires precise points-to analysis.
Similarly, exceptional flows affect points-to analysis, since the path taken by
the exceptional flow can enable or disable object assignments and bindings.

The mutually recursive relationship of CFA and points-to analysis, denoted by edge C,
is obvious: abstract objects (points-to analysis) determine which methods can
be resolved in dynamic dispatch (CFA), while control-flow paths
affect object assignments and bindings for points-to analysis.
In fact, exception-flow analysis is an example of this relationship, which 
exacerbates the edge C relationship further! 
\subsection{Existing approaches}
% WALA appraoch
Existing compilers or analysis frameworks provide a conservative model for exception handling.
One approach assigns all exceptions thrown in a program to a single global variable.
This variable is then read at an exception catch site.
This approach is imprecise since it has no knowledge of 
which exception propagates to a catch site~\cite{shuyingliang:Hendren03scalingjava,shuyingliang:LO2006:Dissertation}.

The second  approach analyzes exceptional control flow only intra-procedurally,
computing only local catch clauses for a try block, 
with no dynamic propagation of exceptions inter-procedurally.%%\footnote{\url{http://sourceforge.net/projects/wala/}}.
% 
%The conservative assumption of exception handling causes  infeasible paths between throw statements and catch blocks.

The third approach is co-analysis using both control-flow analysis and points-to analysis (a.k.a. on-the-fly 
control-flow construction) to handle exceptions, which yields reasonable precision, compared to the aforementioned two 
approaches, as documented in a past precision study~\cite{Bravenboer:2009:Exceptions}.
Unfortunately, even for the best co-analysis, 
where boosting context-sensitivity improves the analysis of
exceptions, it does not improve as much as it does for points-to analysis.
It is too easy for exceptions to cross context boundaries and merge.
For  the previous simple example, we could increase to 1-call-site sensitivity.
 However, context-sensitivity costs more and is easily confused
 when calls are wrapped, as in:
 \begin{center}
 	%% \begin{small}
 	\begin{Verbatim}[fontsize=\relsize{-1}]
 try {
    callsMaybeThrow();  // Call 1
 } catch (Exception e) {  // Handler 1
    System.err.println("Got an exception"); 
 }
 callsMaybeThrow();  // Call 2
 // ...
 void callsMaybeThrow() {
   maybeThrow();
 }
 \end{Verbatim}
 %% \end{small}
 \end{center}
 Similarly, values can easily merge with finitized object-sensitivity in points-to analysis. 
 For example, if object-sensitivity uses $k$ levels of object allocation sites 
 (or a mix with receiver objects) to distinguish contexts,
 objects are merged when the level exceeds $k$.
Even worse, the limited $k$-sensitivity does not distinguish live heap objects
from dead (garbage) heap objects, the existence of which harms both the
precision and performance of the analysis.
  More detailed related work is described in Section~\ref{sec:related}.

\subsection{Our approach}
%Our goal is to provide a systematic abstraction  for the
%analysis of exceptional behavior in object-oriented programs, 
%enabling a spectrum of precision from nullary to precisely matched catches and throws.
%
Due to the intrinsic relationships illustrated in Figure~\ref{fig:tri},
we propose a hybrid joint analysis of pushdown exception-flow analysis with abstract garbage collection enhanced with liveness analysis.
Specifically,  a pushdown system
derived from the concrete semantics of
 a core calculus for an object-oriented language extended with exceptions
is used to tackle exceptional control-flow matching between catches and throws,
 in addition to call and return matches.
Abstract garbage collection is adapted to an object-oriented program setting, 
and it is enhanced with liveness analysis 
to tackle the points-to aspect of exception-flow analysis.
We evaluate an implementation for Dalvik bytecode of the joint analysis technique on 
 a standard set of Java benchmarks.
  The results show that the pruned, pushdown exception-flow analysis  yields higher precision than traditional exception-flow analysis by up to 11 times within a reasonable amount of analysis time. 
\subsection{Organization}
The rest of the paper is organized as follows:
Section~\ref{sec:FJ} presents  the core calculus of an object-oriented language extended with exceptions.
Section~\ref{sec:me-FJ} formulates the concrete semantics for the language with the intent of refactoring and abstracting it into a static analyzer.
Section~\ref{sec:pushdown-sem} derives the abstract semantics from the concrete semantics by reformulating the structure of continuations into a list of frames and forms an implicit pushdown system. 
Section~\ref{subsec:agcoo} introduces the adaptation of  abstract garbage collection in object-oriented languages. 
Section~\ref{subsec:live}  enhances the adapted abstract garbage collection with liveness analysis for better precision.
The reachability algorithm  is described in Section~\ref{sec:reacha}.
Section~\ref{sec:implemenation} describes the details of our implementation.
The evaluation and benchmarks are reported in Section~\ref{sec:evaluation}.
Section~\ref{sec:related} reports related work, and
Section~\ref{sec:conclusion} concludes.

%% file: content2.tex
\section{A Featherweight Java with Exceptions}\label{sec:FJ}

For presentation purpose,
 we start with a variant of Featherweight Java~\cite{shuyingliang:Igarashi:2001:FJM} 
 in ``A-Normal'' form~\cite{Flanagan:1993:ANF}
 with exceptions.
%We start with 
%a variant of Featherweight Java~\cite{shuyingliang:Igarashi:2001:FJM} 
% in ``A-Normal'' form~\cite{Flanagan:1993:ANF}
% with exceptions.
%
A-Normal Featherweight Java (ANFJ) is identical to ordinary Featherweight
Java, except that arguments to a function call must be atomically
evaluable, as they are in A-Normal Form $\lambda$-calculus.
For example, the body {\tt return f.foo(b.bar());}
becomes the sequence of statements 
\begin{center}
\begin{Verbatim}[fontsize=\relsize{-1}]
B b1 = b.bar();
F f1 = f.foo(b1);
return f1;
\end{Verbatim}
\end{center}
This does not change the expressive power of the language or the
nature of the analysis to come, 
but it does simplify the semantics while preserving the essence of the language.

The following grammar describes A-Normal Featherweight Java extended with exceptions;
like regular Java, ANFJ has statement forms:

\vspace{-10pt}
\begin{small}
\begin{align*}
  %\classDef \in
  \syn{Class} &\produces \classform{\className}{\className'}{
    \overrightarrow{\className''\; \fieldName\; \ttsc}\; 
    \constDef\;
    \overrightarrow{\methodDef}
  }
  \\
  \constDef \in \syn{Konst} 
  &
  \produces
  \className\; \ttlp 
   \overrightarrow{
   \className\; \fieldName\;
  }
  \ttrp
  \ttlc
  \mathtt{super}\ttlp 
   \overrightarrow{\fieldName'}
  \ttrp\;
  \ttsc\;
   \overrightarrow{\mathtt{this} \ttfs \fieldName'' \mathrel{=} \fieldName''' \ttsc}
  \ttrc
  \\
  \methodDef \in \syn{Method} 
  &
  \produces
  \className\; \methodName\; \ttlp 
   \overrightarrow{
   \className\; \vv\;
  }
  \ttrp\;
  \ttlc\;
  \overrightarrow{\className\; \vv\; \ttsc}\;
  \vec{\stmt}
  \;\ttrc
  \\
  \stmt \in \syn{Stmt} &
  \produces
  \vv =  \expr \; \ttsc^\lab
  \\
  &\;\;\opor\;\;
  \texttt{return}\; \vv\; \ttsc^\lab
  \\
&\;\;\opor\;\; 
   \texttt{try}\; \{ \vec{s} \}\; 
   \texttt{catch}\; \ttlp C\; \;v \ttrp \;
   \{ \vec{s}' \} ^\lab
%  \texttt{pushhandler}\; \className \; \lab \opor\; 
%  \texttt{pophandler} \opor\; 
\\
&\;\;\opor\;\; \texttt{throw} \; \vv\; \ttsc^\lab
  \\
  \expr \in \syn{Exp} &
  \produces
  \vv 
  \opor 
  \vv \ttfs \fieldName
  \opor
  \vv \ttfs \methodName \ttlp 
   \overrightarrow{\vv}
  \ttrp
  \opor
  \mathtt{new} \; \className\; \ttlp 
   \overrightarrow{\vv}
  \ttrp
  \opor
  \ttlp \className \ttrp \vv 
\end{align*}
\end{small}
\vspace{-8pt}
\begin{small}
\begin{align*}
  \fieldName \in \syn{FieldName} &= \syn{Var}
  \\
  \className \in \syn{ClassName} &\text{ is a set of class names}
  \\
  \methodName \in \syn{MethodCall} &\text{ is a set of method invocation sites}
  \\
  \lab \in \syn{Lab} &\text{ is a set of labels}
  \\
  \vv \in \syn{Var} &\text{ is a set of variables}
\end{align*}
\end{small}%
The set $\syn{Var}$ contains both variable and field names.
Every statement has a label.
%
% The language includes labels on statements to provide a
% convenient way of referencing program points.
%
The function $\ssucc : \syn{Lab} \parto \syn{Stmt}$ yields the
(semantically) subsequent statement for a statement's label.

\section{Machine semantics for Featherweight Java}\label{sec:me-FJ}

\setlength{\abovecaptionskip}{-5pt}
\begin{figure}
{\small
		\begin{align*}
		\state \in \State &= 
		\syn{Stmt} \times 
		\FramePointer \times 
		\s{Store} \times
		\s{Kont} \times
		\s{Time}  
		\\
		\store \in \s{Store} &= \s{Addr} \parto \s{D}
		\\
		\den \in \Den &= \s{Val}
		\\
		\val \in \s{Val} &= \s{Obj} %+ \s{Kont}
		\\
		\obj \in \s{Obj} &= \syn{ClassName} \times\ObjectPointer
		\\
		\cont \in \s{Kont} &\produces
		%   \syn{Var} \times \syn{Stmt} \times \s{BEnv} \times \s{KontAddr}
		\mathbf{fun}(\vv,\stmt,\fp,\cont)
		\\
		&\;\;\opor\;\;
		%   \syn{Var} \times \syn{Stmt} \times \s{BEnv} \times \s{KontAddr}
		\mathbf{handle}(\className,\vv,\vec{\stmt},\fp,\cont)
		\\    
		&\;\;\opor\;\; 
		\mathbf{halt}
		\\
		\fp \in &~\FramePointer  \text{ is a set of frame pointers }
		\\
		\objp \in &~\ObjectPointer  \text{ is a set of object pointers}
		\\
		\mathit{ptr} \in \mathit{Ptr} &= \FramePointer + \ObjectPointer
		\\
		\addr \in \s{Addr} &= (\syn{Var} + \syn{Method}) \times\mathit{Ptr}
		%  \\
		%  \contptr \in \s{KontPtr} &\subseteq \s{Addr}
		\\
		\tm \in \s{Time} &\text{ is a set of time-stamps}  
		\text.  
		\end{align*}
	%% \end{small}%
}
	\caption{Concrete state-space for A-Normal Featherweight Java.}
	\label{fig:java-concrete-state-space}
\end{figure}

\begin{figure}
  {\small
  \begin{align*}
    \FetchRuctor &: \syn{ClassName} \to (\syn{FieldName}^* \times \s{Ructor})
    \\
  \Ructor &\in 
  \s{Ructor} = 
  \overbrace{\s{Addr}^*}^{\text{fields}} 
  \times 
  \overbrace{\Den^*}^{\text{arguments}}
  \to
  (
  \overbrace{
    %(\s{Addr} \to \Den) 
    \s{Store}
  }^{\text{field values}}
  \times 
  \overbrace{
    %(\syn{FieldName} \to \s{Addr})
   \ObjectPointer
  }^{\text{record}}
  )
  \\
  \MethodLookup & : \s{D} \times \syn{MethodCall} \parto \syn{Method}
  \end{align*}
  }
  \caption{Helper functions for the concrete semantics.}
  \label{fig:concrete-anfw-java-helper}
\end{figure}
\setlength{\abovecaptionskip}{3pt}

In preparation for synthesizing an abstract interpreter, 
we first construct a small-step abstract machine-based semantics
for Featherweight Java. 
Figure~\ref{fig:java-concrete-state-space} contains the concrete
state-space for the small-step Featherweight Java machine.
Each machine state has five components: a statement,
a frame pointer, a store, a continuation and a timestamp.
The encoding of objects abstracts over a low-level implementation:
an object is a class plus a base pointer, and field addresses are ``offsets'' from
this base pointer.
Given an object $(\className,\objp)$, the address of field $f$ would be
$(f,\objp)$.
In the semantics, object allocation creates a single new base object pointer $\objp'$.

The concrete semantics use the helper functions described in
Figure~\ref{fig:concrete-anfw-java-helper}.
The constructor-lookup function $\FetchRuctor$ yields the field names
and the constructor associated with a class name.
A constructor $\Ructor$ takes a newly allocated address to use for
fields and a vector of arguments; it returns the change to the store
plus the record component of the object that results from running the
constructor.
The method-lookup function $\MethodLookup$ takes a method invocation point and an
object to determine which method is actually being called at that
point.
The concrete semantics are encoded as a small-step transition relation,
$(\To) \subseteq \State \times \State$.
Each statement and expression type has a transition rule below. \\
% \begin{figure}
\paragraph{Variable reference} Variable reference computes the address relative to the current frame
pointer and retrieves the result from the store:

\vspace{-10pt}
{\small
\begin{gather*}
  (
  \sembr{\vv = \vv' \ttsc^\lab},
  \fp,
  \store,
  \cont,
  \tm
  )
  \To
  (
  \ssucc(\lab),
  \fp,
  \store',
  \cont,
  \tm'
  ),
  \\
  \text{ where } \tm' = \tick(\lab, \tm), ~
    \store' = \store[(\vv, \fp) \mapsto \store(\vv', \fp)]
%\begin{small}\begin{align*}
%  \tm' = \tick(\lab, \tm), ~
%  \store' = \store[(\vv, \fp) \mapsto \store(\vv', \fp)]
%  \text.
%\end{align*}\end{small}%
\text{.}
\end{gather*}
}%
\paragraph{Return to call}
Returning from a function  checks if the top-most frame pointer
is a function continuation (as apposed to an exception-handler continuation).
If it is, then the machine binds the result and restores the context of the
continuation; if not, then the machine skips to the next continuation.
If $\cont = \mathbf{fun}(\vv',\stmt, \fp', \cont')$:

\vspace{-10pt}
{\small
\begin{gather*}
  (\sembr{\texttt{return}\; \vv\; \ttsc^\lab}, \fp, \store, 
   \cont,
   \tm)
  \To
  (\stmt, \fp', \store', \cont', \tm')\text{, where } 
  \\
  \tm' = \tick(\lab, \tm),~\store' = \store[(\vv',\fp) \mapsto \den],~
    \den = \store(\vv,\fp)
%\begin{small}\begin{align*}
%  \store' = \store[(\vv',\fp) \mapsto \den],~
%  \den = \store(\vv,\fp)
%  \text.
%\end{align*}\end{small}%
\text{.}
\end{gather*}
}%
\paragraph{Return over handler}
If the topmost continuation is a handler,
then the machine pops the handler off the stack.
So, if $\cont = \mathbf{handle}(\className,\vv,\vec{\stmt},\fp',\cont')$:

\vspace{-10pt}
{\small
\begin{gather*}
  (\sembr{\texttt{return}\; \vv\; \ttsc^\lab}, \fp, \store, 
   \cont,
   \tm)
  \To
  (\sembr{\texttt{return}\; \vv\; \ttsc^\lab}, \fp, \store, 
   \cont',
   \tm')
   \\
%  (\stmt, \fp', \store', \cont', \tm')\text{, where }
\text{ where }  \tm' = \tick(\lab, \tm)
%\begin{small}
%  \tm' = \tick(\lab, \tm)
%  \text{.}
%\end{small}
%  \\
%\begin{small}\begin{align*}
%  \tm' &= \tick(\lab, \tm)
%  \text.
%\end{align*}\end{small}%
\text{.}
\end{gather*}
}%
\paragraph{Field reference}
Field reference is similar to variable reference, except that it
must find the base object pointer with which to compute
the appropriate offset:

\vspace{-10pt}
{\small
\begin{gather*}
  (
  \sembr{\vv = \vv'\ttfs \fieldName\; \ttsc^\lab},
  \fp,
  \store,
  \cont,
  \tm
  )
  \To
  (
  \ssucc(\lab),
  \fp,
  \store',
  \cont,
  \tm'
  ),
  \\
  \text{ where }  \tm'  = \tick(\lab, \tm),
  \\
    (\className, \op') = \store(\vv',\fp),~
    \store' = \store[(\vv,\fp) \mapsto \store(\fieldName,\op')]
%\begin{aligned}
%  \tm' & = \tick(\lab, \tm)
%  \\
%  (\className, \op') &= \store(\vv',\fp)
%  \\
%  \store' &= \store[(\vv,\fp) \mapsto \store(\fieldName,\op')]
  \text.
%\end{aligned}
\end{gather*}
}%
\paragraph{Method invocation}
Method invocation is a multi-step process:
it looks up the object,  determines the class of the object 
and then identifies the appropriate method. 
When transitioning to the body of the resolved method, 
a new function continuation is instantiated, which 
records the caller's execution context. Finally, the
store is updated with the bindings of formal parameters to
evaluated values of passed arguments.

\vspace{-10pt}
{\small
\begin{gather*}
\begin{split}
  &(
  \sembr{\vv = \vv_0 \ttfs \methodName \ttlp
    \overrightarrow{\vv'}
    %\vv_1 \ttcm \ldots \ttcm \vv_n
  \ttrp \ttsc^\lab},
  \fp,
  \store,
  \cont,
  \tm
  )
  \To
  (
  \stmt_0,
  \fp',
  \store',
  \cont',
  \tm'
  ),
\end{split}
\\
\text{ where }
  \methodDef = 
  \sembr{\className\; \methodName\; \ttlp 
   \overrightarrow{
   \className\; \vv''\;
  }
  \ttrp\;
  \ttlc
  \overrightarrow{\className'\; \vv'''\; \ttsc}\;
  \vec{\stmt}
  \ttrc}
  = \MethodLookup(\den_0,\methodName)\\
%% \begin{align*}
  \den_0 = \store(\vv_0,\fp),
  \den_i = \store(\vv'_i,\fp),
  \tm' = \tick(\lab,\tm),
    \fp' = \alloc(\lab,\tm'),\\
  \cont' = \mathbf{fun}(\vv,\ssucc(\lab), \fp, \cont),
%  \contptr' &= \alloc_\cont(\methodDef,\tm')
%  &
  \addr'_i = (\vv_i'',\fp'),
%  \\
%  \\
%  \addr''_j &= (\vv_j''',\fp')
%  &
%  \benv' &= [\sembr{\tt this} \mapsto \vv_0)]
%  \\
%  \benv'' &= \benv'[\vv_i'' \mapsto \addr_i', \vv'''_j \mapsto \addr_j'']
%  &
  \store' = \store [\addr'_i \mapsto \den_i]
  \text.
%% \end{align*}
\end{gather*}}%
\paragraph{Object allocation}
Object allocation creates a new base object pointer;
it also invokes the constructor helper to initialize the object(
The $(+)$ operation represents right-biased functional
union in that wherever vector $\vec{x}$ is in scope, its
components are implicitly in scope: $\vec{x} = \vect{x_0, \ldots, x_{\mathit{length(\vec{x})}}}$):

\vspace{-10pt}
{\small
\begin{gather*}
\begin{split}
  &(
  \sembr{
    \vv = 
    {\tt new}\; \className\; \ttlp
    %\vv_1 \ttcm \ldots \ttcm \vv_n
    \overrightarrow{\vv'}
  \ttrp \ttsc^\lab},
  \fp,
  \store,
  \cont,
  \tm
  )
  \To
  (
  \ssucc(\lab),
  \fp,
  \store',
  \cont,
  \tm'
  )\text{,}
\\
\end{split} \\
\text{ where } \tm' = \tick(\lab,\tm),~~
%% \begin{small}\begin{align*}
  \den_i = \store(\vv_i',\fp),~~
  (\vec{\fieldName},\Ructor) =
   \FetchRuctor(\className),~~ \\
  \fp' = \alloc(\lab,\tm'),~~
  \addr_i = (\fieldName_i,\fp')
  (\Delta \store, \op') = 
   \Ructor(\vec{\addr}, \vec{\den}),~~ \\
  \den' = (\className, \op'),~~
  \store' = \store + \Delta \store + [(\vv,\fp) \mapsto \den']
  \text.
%% \end{align*}\end{small}%
\end{gather*}
}%
\paragraph{Casting}
A cast references a variable, replacing the class of the object:

\vspace{-10pt}
{\small
\begin{gather*}
  (
  \sembr{\vv = \ttlp C'\ttrp\; \vv'},
  \fp,
  \store,
  \cont,
  \tm
  )
  \To
  (
  %\dot \vexpr[\den],
  \ssucc(\lab),
  \fp,
  \store',
  \cont,
  \tm'
  ),
  \\
  \text{ where }
   \tm' = \tick(\lab, \tm),~~
    \store' = \store[(\vv,\fp) \mapsto \store(\vv',\fp)]
%\begin{small}\begin{align*}
%  \tm' = \tick(\lab, \tm)
%  \store' = \store[(\vv,\fp) \mapsto \store(\vv',\fp)]
%  \text.
%\end{align*}\end{small}%
\text{.}
\end{gather*}
}%
\paragraph{Try}
A {\tt try} statement creates a new handler continuation and then
proceeds to the body of the {\tt try} statement.

\vspace{-10pt}
{\small
\begin{gather*}
  (\sembr{
   \texttt{try}\; \{ \vec{s} \}\; 
   \texttt{catch}\; \ttlp C\; \;v \ttrp \;
   \{ \vec{s}' \} ^\lab
    }
    , \fp, \store, \cont, \tm) \\
  \To
  (\ssucc(\lab), \fp, \store, \cont', \tm') \\
  \text{ where }
   \tm' = \tick(\lab, \tm),~~
    \cont' =  \mathbf{handle}(\className,\vv',\stmt'_1, \fp, \cont)
%\begin{small}\begin{align*}
%  \tm' &= \tick(\lab, \tm)
%  \\
%  \cont' &=  \mathbf{handle}(\className,\vv',\stmt'_1, \fp, \cont)
% % \\
% % \den &= \store(\benv(\vv))
% % &
%%  \store' &= \store[\benv'(\vv') \mapsto \den]
%  \text.
%\end{align*}\end{small}%
\text{.}
\end{gather*}
}%
\paragraph{Throw to matching handler}
%
% To simplify throwing an exception, we're going to slightly extend the syntax of
% the language to allow denotable values to be thrown directly within the syntax:
% \begin{align*}
%  \syn{Stmt} &\produces \cdots \opor \texttt{throw}\; \den\ttsc^\lab
% \end{align*}
% When we encounter a syntactic \texttt{throw} statement, 
% we convert it to one of these semantic throw statements:
% If $\cont = \mathbf{fun}(\vv',\stmt, \fp', \cont')$:
% \begin{gather*}
%   (\sembr{\texttt{throw}\; \vv\; \ttsc^\lab}, \fp, \store, \cont, \tm)
%   \To
%   (\sembr{\texttt{throw}\; \den\; \ttsc^\lab}, \fp, \store, \cont, \tm')
%   \text{, where }
%   \\
% \begin{small}\begin{align*}
%   \tm' &= \tick(\lab, \tm)
%   &
%   \den &= \store(\vv,\fp)
%   \text.
% \end{align*}\end{small}%
% \end{gather*}
When the machine encounters a throw statement, it must check if
the topmost continuation is both a handler and a matching handler;
if so, then it returns to the context within the continuation:
If $\cont = \mathbf{handle}(\className',\vv',\stmt, \fp'', \cont')$
and $\store(\vv,\fp) = (\className,\op')$
and $\className$ is a $\className'$:

\vspace{-10pt}
{\small
\begin{gather*}
  (\sembr{\texttt{throw}\; \vv\; \ttsc^\lab}, 
     \fp, \store, \cont, \tm)
  \To
  (\stmt, \fp'', \store', \cont', \tm') \\\text{where~~  }
  %% \begin{small}\begin{align*}
   \tm' = \tick(\lab, \tm),~~
  \store' = \store[(\vv',\fp'') \mapsto (\className,\op')]
  \text.
%% \end{align*}\end{small}%
\end{gather*}}%
\paragraph{Throw past non-matching handler}
When throwing, if the topmost handler is not a match,
the machine looks deeper in the stack for a matching handler.
If $\cont = \mathbf{handle}(\className',\vv',\stmt, \fp'', \cont')$
and $\store(\vv,\fp) = (\className,\op')$
but $\className$ is not a $\className'$:

\vspace{-10pt}
{\small
\begin{gather*}
  (\sembr{\texttt{throw}\; \vv\; \ttsc^\lab}, \fp, \store, \cont, \tm)
  \To
 (\sembr{\texttt{throw}\; \vv\; \ttsc^\lab}, \fp, \store, \cont', \tm')
 \\ \text{where ~~ }   \tm' = \tick(\lab, \tm)
%  \\
%\begin{small}\begin{align*}
%  \tm' &= \tick(\lab, \tm)
%  \text.
%\end{align*}\end{small}%
\text{.}
\end{gather*}}%
\paragraph{Throw past return point}
If throwing an exception and the topmost handler is a function return point,
then it jumps over this continuation.
If $\cont = \mathbf{fun}(\vv',\stmt,\fp',\cont')$:

\vspace{-10pt}
{\small
\begin{gather*}
  (\sembr{\texttt{throw}\; \vv\; \ttsc^\lab}, \fp, \store, \cont, \tm)
  \To
 (\sembr{\texttt{throw}\; \vv\; \ttsc^\lab}, \fp, \store, \cont', \tm')
 \\ \text{where ~~ }   \tm' = \tick(\lab, \tm)
%  \\
%\begin{small}\begin{align*}
%  \tm' &= \tick(\lab, \tm)
%  \text.
%\end{align*}\end{small}%
\text{.}
\end{gather*}}%
\paragraph{Popping handlers}
When control passes out of a {\tt try} block,
the topmost handler must be popped from the stack.
To handle this, the ``successor'' of  the last 
statement in a {\tt try} block is actually a
special {\tt pophandler} statement,
and the ``successor'' of that statement is the statement
directly following the {\tt try} block.

\vspace{-10pt}
{\small
\begin{gather*}
 (\sembr{\texttt{pophandler}\; \ttsc^\lab}, \fp, \store, \cont, \tm)
  \To
 (\mathit{succ}(\lab), \fp, \store, \cont', \tm')
 \\ \text{ where ~~ }
  \tm' = \tick(\lab, \tm),~~
    \cont = \mathbf{handle}(\ldots,\cont')
%\begin{small}\begin{align*}
%  \tm' = \tick(\lab, \tm),~~
%  \cont = \mathbf{handle}(\ldots,\cont')
%  \text.
%\end{align*}\end{small}%
\text{.}
\end{gather*}}%
\vspace{-16pt}

\section{A pushdown semantics of exceptions}\label{sec:pushdown-sem}
\setlength{\abovecaptionskip}{-5pt}
\begin{figure}
{\small
  \begin{align*}
  \astate \in \aState = 
  \syn{Stmt} & \times
  \aFramePointer \times 
  \sa{Store} \times
  \sa{Kont} \times
  \sa{Time}  
  \\
  \astore \in \sa{Store} &= \sa{Addr} \parto \sa{D}
  \\
  \aden \in \aDen &= \Pow{\sa{Val}}
  \\
  \aval \in \sa{Val} &= \sa{Obj} %+ \sa{Kont} 
  \\
  \aobj \in \sa{Obj} &= \syn{ClassName} \times \aObjectPointer
  \\
  \acont \in \sa{Kont} &= \sa{Frame^*}
  \\
  \aphrame \in \sa{Frame} & =  \sa{CallFrame}  + \sa{HandlerFrame} 
  \\ 
  \acallframe \in \sa{CallFrame}  &::=    \mathbf{fun}(\vv,\stmt,\afp)
  \\
  \aexnframe \in \sa{HandlerFrame} &:: =   \mathbf{handle}(\className,\vv,\vec{\stmt},\afp)
 % \\
 % \acont \in \sa{Kont} &\produces
%   \syn{Var} \times \syn{Stmt} \times \s{BEnv} \times \s{KontAddr}
 %   \mathbf{fun}(\vv,\stmt,\afp,\acont)
%
 %  \\
  % &\;\;\opor\;\;
%   \syn{Var} \times \syn{Stmt} \times \s{BEnv} \times \s{KontAddr}
 %   \mathbf{handle}(\className,\vv,\vec{\stmt},\afp,\acont)
%
%  \\    
 % &\;\;\opor\;\; 
 %   \mathbf{halt}
  \\
 \afp \in \aFramePointer & \text{ is a set of frame pointers }
 \\
 \aobjp \in \aObjectPointer & \text{ is a set of object pointers } 
  \\
    \widehat{\mathit{ptr}} \in \widehat{\mathit{Ptr}}&= \aFramePointer + \aObjectPointer
    \\
  \aaddr \in \sa{Addr} &= (\syn{Var} + \syn{Method}) \times \widehat{\mathit{Ptr}}%\aFramePointer 
  \\
  \atm \in \sa{Time} &\text{ is a set of time-stamps}  
  \text.  
\end{align*}  }
\caption{Abstract state-space for pushdown analysis of A-Normal Featherweight Java.}
\label{fig:java-abstract-state-space-pushdown}
\end{figure}
\setlength{\abovecaptionskip}{5pt}

With the concrete semantics for A-Normal Featherweight Java with exceptions in place, 
we are ready to derive the abstract semantics for static analysis.
``Abstracting abstract machines'' (AAM) has proposed a systematic approach to derive such kind of abstraction, 
which is equivalent to most of  the conservative static analyses~\cite{shuyingliang:VanHorn:2010:Abstract}.
The idea is to make the analysis finite and terminate by finitize every component in the state,
so that there is no source of infinity. 
However, when we apply this technique, the precision is not satisfiable in the client security analysis~\cite{shuyingliang:LiangMH13},
% ~\cite{shuyingliang:LiangMH13},
because the over-approximation of the continuation component causes spurious control-flow and return-flows.

Therefore, in this work, we choose to abstract \emph{less} than what AAM approach does: 
we leave the stack (represented as continuation) unbounded in height.
In fact, the central idea behind this abstraction is the generalization of two kinds of frames on stack:
the function frames and the exception-handler frames.
In this way, we form the abstract pushdown semantics. 
Then, the pushdown abstract semantics will further be computed as control-state reachability in pushdown systems,
which is  evolved from the 
work of~\cite{mattmight:Reps:1998:CFL,shuyingliang:Earl:2010:Pushdown,shuyingliang:Earl:2012:IPDCFA}.
However, unlike them, we improve the algorithm to handle  new behaviors introduced by exceptions.
The algorithm is detailed in Section~\ref{sec:reacha}.
The rest of the section focuses how we formulate the pushdown semantics.
%Fortunately, on this semantics, 
%we can still compute the reachable states via an enhanced reachability algorithm, which is evolved from the 
%work of Reps~\cite{mattmight:Reps:1998:CFL} and Earl~\etal{}~\cite{shuyingliang:Earl:2010:Pushdown,shuyingliang:Earl:2012:IPDCFA}.
%However, unlike previous work, we  extend control-state reachability in pushdown systems to handle the new behaviors introduced by exceptions.
%The algorithm is detailed in Section~\ref{sec:reacha}.
%The rest of the section focuses how we formulate the pushdown semantics.
%%to yield a finite state-based analysis, 
%which is equivalent to most of  the conservative static analyses.

%Rather than adopting the technique of 
%``Abstracting abstract machines'' (AAM)
%to yield a finite state-based analysis, 
%which is equivalent to most of  the conservative static analyses,
%we choose to lightly reformulate   the concrete semantics
%and conduct a similar structural abstraction---pushdown analysis.
%%
%This approach abstracts \emph{less} than AAM does: 
%we leave the stack unbounded in height.
%%
%Ultimately, we  will  extend  control-state reachability in pushdown systems to handle the new behaviors introduced by exceptions.
Abstract semantics are defined on an abstract state-space.
To formulate  the pushdown abstract state-space,
we first reformulate continuations as a list of frames in the concrete semantics:
%% \begin{small}

\vspace{-10pt}
{\small
\begin{align*}
 \s{Kont} &\cong \s{Frame}^*
 \\
 \s{Frame} & =  \s{CallFrame}  + \s{HandlerFrame} 
 \\ 
 \s{CallFrame}  &::=    \mathbf{fun}(\vv,\stmt,\fp)
 \\
 \s{HandlerFrame} &:: =   \mathbf{handle}(\className,\vv,\vec{\stmt},\fp)
 \text.
\end{align*}
}%
%% \end{small}
We have two kinds of frames: function frames as well as handler frames.
As with continuations, they may grow without bound (The enhanced reachability algorithm handles this in Section~\ref{sec:reacha}).
%
%Fortunately, we can still compute the reachable states via an enhanced reachability algorithm, which is evolved from the 
%work of Reps~\cite{mattmight:Reps:1998:CFL} and Earl~\etal{}~\cite{shuyingliang:Earl:2010:Pushdown,shuyingliang:Earl:2012:IPDCFA}.
%The algorithm is detailed in Section~\ref{sec:reacha}.

Figure~\ref{fig:java-abstract-state-space-pushdown} contains the abstract
state-space for the pushdown version of the small-step Featherweight Java machine.
At this point, we can extract the high-level structure of the pushdown system from the state-space.
A configuration in a pushdown system is a control state (from a finite set) paired with a stack (with a finite number of frames that are defined in Figure~\ref{fig:java-abstract-state-space-pushdown}). This can be observed as follows:
%
%% \begin{small}

\vspace{-10pt}
{\small
\begin{align*}
 \aState &=
  \syn{Stmt} \times 
  \aFramePointer \times 
  \sa{Store} \times
  \sa{Kont} \times
  \sa{Time} 
  \\
  &\cong
  \syn{Stmt} \times 
  \aFramePointer \times 
  \sa{Store} \times
  \sa{Time} \times 
  \sa{Kont}
  \\
  &=
  \left(\syn{Stmt} \times 
 \aFramePointer \times 
  \sa{Store} \times
  \sa{Time}\right) \times 
  \sa{Kont} 
  \\
  &=
  \underbrace{
  \left(\syn{Stmt} \times 
  \aFramePointer \times 
  \sa{Store} \times
  \sa{Time}\right)}_{\text{control states}}
  \times 
  \underbrace{\sa{Frame}^*}_{\text{stack}}
\end{align*} 
%% \end{small}
}%

Now let us show the detailed abstract transition relations.
%\footnote{The $\join$ operation is a weak update operator in the abstract semantics.}.
Thanks to the way we do the abstraction so far (That is, structural abstraction of concrete states except for the stack component), 
the abstract transition relations resemble a lot as their concrete counterparts. 
The biggest difference in abstract semantics is that it does weak updates using the operator $\join$.
For example, for variable reference (weak updates are underlined.):

\vspace{-10pt}
{\small
\begin{gather*}
  (
  \sembr{\vv = \vv' \ttsc^\lab},
  \afp,
  \astore,
  \acont,
  \atm
  )
  \aTo
  (
  \ssucc(\lab),
  \afp,
  \astore',
  \acont,
  \atm'
  )
  \text{,}\\ \text{ where } \atm' = \atick(\lab, \atm) \text{ and }
  \underline{\astore' = \astore \join [(\vv, \afp) \mapsto \store(\vv', \afp)]}
 %% \text{[Variable reference]}\text{.}
\end{gather*}
}
The other difference is, whenever evaluating expressions, the results are abstract entities that represents one or more concrete entities. For example,  field reference:
{\small
\begin{gather*}
  (
  \sembr{\vv = \vv'\ttfs \fieldName\; \ttsc^\lab},
  \afp,
  \astore,
  \acont,
  \atm
  )
  \aTo
  (
  \ssucc(\lab),
  \afp,
  \astore',
  \acont,
  \atm'
  )
  \\
  \text{ where }
  \atm' = \atick(\lab, \atm),
    \underline{  (\className, \aobjp') \in \astore(\vv',\afp)},
    \astore' = \astore \join [(\vv,\afp) \mapsto \astore(\fieldName,\aobjp')]
\end{gather*}
}%
The underlined operation shows that there could be more than one abstract objects are evaluated.
The two differences apply to all the other rules. 
To save space, we demonstrate the abstract rules that involve exceptions. 
%as shown in Fig~\ref{fig:pushdown-semantic-exception}.
%In 
Fig~\ref{fig:pushdown-semantic-exception} shows 
how we handle the exception-flow and its mix with normal control-flow.
The idea is the ``multi-pop'' behavior introduced when a function call returns or an exception throws (as the concrete semantics).
The effect of this approach substantially simplifies the control-reachability algorithm during summarization,
as we shall show in Section~\ref{sec:reacha}.

\begin{figure}
\fbox{
%% \begin{small}
  \small
\begin{minipage}[b]{0.85\linewidth}
\text{[Try]}:
\begin{gather*}
  (\sembr{
   \texttt{try}\; \{ \vec{s} \}\; 
   \texttt{catch}\; \ttlp C\; \;v \ttrp \;
   \{ \vec{s}' \} ^\lab
    }
    , \afp, \astore, \acont, \atm) \\
  \aTo
  (\ssucc(\lab), \afp, \astore, \acont', \atm'), \\
  \text{ where } 
%\begin{small}\begin{align*}
  \atm' = \atick(\lab, \atm)
 \text{ and }  
  \acont' =  \mathbf{handle}(\className,\vv',\stmt'_1, \afp) :: \acont
  \text{.}
\end{gather*}

\text{[Throw to matching handler]:}
\begin{gather*}
  (\sembr{\texttt{throw}\; \vv\; \ttsc^\lab}, 
     \afp, \astore, \acont, \atm)
  \aTo
  (\stmt, \afp'', \astore', \acont', \atm')\text{,} \\
  \text{ where }
   \atm' = \atick(\lab, \atm),
  \astore'= \astore \join [(\vv',\afp'') \mapsto (\className,\aobjp')], \\
  \acont = \mathbf{handle}(\className',\vv',\stmt, \afp'') :: \acont',
  (\className,\aobjp') \in \astore(\vv,\afp) \text{,}
  \className \text{ is a } \className'
  \text{.}
\end{gather*}

\text{[Throw past non-matching handler]:}
\begin{gather*}
  (\sembr{\texttt{throw}\; \vv\; \ttsc^\lab}, \afp, \astore, \acont, \atm)
  \aTo
 (\sembr{\texttt{throw}\; \vv\; \ttsc^\lab}, \afp, \astore, \acont', \atm')
 \text{,} \\
 \text{ where }
  \atm' = \atick(\lab, \atm),
  \acont = \mathbf{handle}(\className',\vv',\stmt, \afp'') :: \acont', \\
  (\className,\aobjp') \in \astore(\vv,\afp) \text{ and }
  \className \text{ is not  a } \className'
  \text{.}
\end{gather*}

 \text{[Throw past return point]:} 
\begin{gather*}
  (\sembr{\texttt{throw}\; \vv\; \ttsc^\lab}, \afp, \astore, \acont, \atm)
  \aTo
 (\sembr{\texttt{throw}\; \vv\; \ttsc^\lab}, \afp, \astore, \acont', \atm')
 \text{,} \\
 \text{ where }
  \atm' = \atick(\lab, \atm),
\text{ and } \acont = \syn{fun}(\vv',\stmt,\afp') \text{::} \acont'
\text{.}
\end{gather*}

\text{[Return over handler]:}
\begin{gather*}
  (\sembr{\texttt{return}\; \vv\; \ttsc^\lab}, \afp, \astore, 
   \acont,
   \atm)
  \aTo
  (\sembr{\texttt{return}\; \vv\; \ttsc^\lab}, \afp, \astore, 
   \acont',
   \atm'), \\
\text{ where } \atm' = \atick(\lab, \atm)
\text{ and } \acont = \text{handle}(\className,\vv,\vec{\stmt},\afp') ::  \acont'
\text{.}
\end{gather*}

\text{[Popping handlers]:}
\begin{gather*}
 (\sembr{\texttt{pophandler}\; \ttsc^\lab}, \afp, \astore, \acont, \atm)
  \aTo
 (\mathit{succ}(\lab), \afp, \astore, \acont', \atm'), \\
 \text{ where }
  \atm' = \atick(\lab, \atm)   
  \text{ and }
  \acont = \mathbf{handle}(\ldots) :: \acont'
  \text{.}
\end{gather*}
\end{minipage}
%% \end{small}
}
\caption{Abstract transition relations (exception)}
\label{fig:pushdown-semantic-exception}
\end{figure}

\section{Enhanced abstract garbage collection}\label{sec:eagc}

The previous section formulates a pushdown system to handle complicated control-flows (both normal and exceptional). 
This section describes how we prune the analysis for exceptions from the angle of  points-to analysis  with enhanced garbage collection generalized for object-oriented programs.

\subsection{Abstract garbage collection in an object-oriented setting}\label{subsec:agcoo}
The idea of abstract garbage collection was first 
proposed in the work of Might and Shivers~\cite{Might:2006:GammaCFA} for higher-order programs.
As an analog to the concrete garbage collection, 
abstract garbage collection reallocates unreachable abstract resources.
Order-of-magnitude improvements in precision have been reported, 
even as it drops run-times by cutting away false positives.
It is natural to think that this technique can benefit 
exception-flow analysis for object-oriented languages.
In fact, in an object-oriented setting, 
abstract garbage collection can free
 the analysis from the context-sensitivity and object-sensitivity limitation,
 since the ``garbage'' discarded is ignorant of any form of sensitivity!
For example, in the following simple code snippet,
\begin{center}
%{\fontsize{9pt}{4pt}\selectfont
\begin{Verbatim}[fontsize=\relsize{-1}]
  A a1 = idA(new A());
  A a2 = idA(new A()):
  B b1 = idB(a1.makeB());
  B b2 = idB(a2.makeB());
  \end{Verbatim}
%}
\end{center}
\texttt{idA} and \texttt{idB} are identity functions.
Traditionally, with one level of object-sensitivity and  one level of context sensitivity, 
we are able to distinguish the arguments passed in all of the four lines.
However,
it is easy to exceed the $k$-sensitivity 
(call site, allocation sites, receiver objects, \etc) in modern software constructs.
Abstract garbage collection can play a role  in the way that 
it  discards conservative values 
and enables fresh bindings for reused variables (formal parameters).
This does not need knowledge about any sensitivity! 
Thus, it  can avoid ``merging'' of abstract object values 
(and so indirectly eliminate potentially spurious function calls).
For exceptions specifically, abstract garbage collection can 
help avoid conflating exception objects 
at various throw sites.

To gain the promised analysis precision and performance, 
we must  conduct a careful and subtle
    redesign of the abstract garbage collection machinery 
    for object-oriented languages.
Specifically, we need to make it work 
    with the abstract semantics defined in Section~\ref{sec:pushdown-sem}.
    In addition, the reachability algorithm should also be able to work with abstract garbage collection.  
    Fortunately, the challenge of how to adapt abstract garbage collection into pushdown systems has been resolved in the work of Earl~\etal{}~\cite{shuyingliang:Earl:2012:IPDCFA}. 
    Here we focus on the enhanced machinery for object-oriented languages.
    %We will review this work in Section~\ref{subsec:setup}.
    
First, we describe how we 
adapt abstract garbage collection to analyze object-oriented languages.
Abstract garbage collection discards unreachable elements from the
store, 
it modifies the transition relation
to conduct a ``stop-and-copy'' garbage collection before each
transition.
To do so, we define a garbage collection function 
%$\aCollect : \sa{Conf} \to \sa{Conf}$
$\aCollect : \aState\to \aState$
on
configurations:

\vspace{-10pt}
{\small
\begin{align*}
\aCollect(\overbrace{\vec{s},\afp,\astore,\acont, \atm}^{\astate})
&= (\vec{s},\afp,\astore|\mathit{Reachable}(\astate),\acont)
\text,
\end{align*} }%
%\begin{align*}
%\aCollect(\overbrace{\vec{s},\afp,\astore,\acont}^{\aconf})
%&= (\vec{s},\afp,\astore|\mathit{Reachable}(\aconf),\acont)
%  \text,
%  \end{align*}
  where the pipe operation $f|S$ yields the function $f$, but with
  inputs not in the set $S$ mapped to bottom---the empty set.
  The reachability function $\mathit{Reachable} :\aState \to \PowSm{\sa{Addr}}$ %\sa{Conf} \to \PowSm{\sa{Addr}}$
  first computes the root set and then the transitive closure of an
  address-to-address adjacency relation: 

  \vspace{-10pt}
  {\small
  \begin{align*}
  \mathit{Reachable}(\overbrace{\vec{s},\afp,\astore,\acont, \atm}^\astate) &=
\setbuild{ \aaddr }{ \aaddr_0 \in \mathit{Root}(\astate)
  \text{ and }
  \aaddr_0  
    \mathrel{\areaches_\astore^*}
  \aaddr
}
\text,
  \end{align*} }%
where the function $\mathit{Root} : \aState \to  \PowSm{\sa{Addr}}$ finds the root addresses:

\vspace{-10pt}
  {\small
  \begin{gather*} 
  \mathit{Root}(\vec{s}, \afp,\astore,\acont, \atm)  = \{( v,\afp): (v, \afp) \in \mathit{dom}(\astore) \} \union
\mathit{StackRoot}(\acont)
  \end{gather*} } %
The $\mathit{StackRoot} : \sa{Kont} \to \PowSm{\sa{Addr}}$ function
  finds roots on the stack. 
However, only $\aCallFrame$ has the component to construct addresses, so we define a 
helper function $\hat{\mathcal{F}} : \sa{Kont} \to \aCallFrame^*$ 
to extract only $\aCallFrame$ out from the stack and skip over all the handle frames.
Now $\mathit{StackRoot} $ is defined as

\vspace{-10pt}
{\small
 \begin{gather*} 
   \mathit{StackRoot}(\acont)  = 
 \{(v, \afp_i) : (v, \afp_i) \in \mathit{dom}(\astore) ~\text{and}~ \afp_i \in \hat{\mathcal{F}}(\acont)\}
  \text,
  \end{gather*}
}%
and the relation:$ (\areaches) \subseteq \sa{Addr} \times \sa{Store} \times \sa{Addr} $ connects adjacent addresses: $ \aaddr \mathrel{\areaches_\astore}  \aaddr' \text{ iff there exists }(\className, \aobjp) \in \astore(\aaddr)$
such that 
  $\aaddr' \in \{( \fieldName, \aobjp): (\fieldName, \aobjp  ) \in \mathit{dom}(\astore)\}$.
The formulated  abstract garbage collection semantics constructs the subroutine \texttt{eagc} that is called  in Alg.~\ref{algo:stepIPDS}, which is the interface to enable abstract garbage collection in the reachability algorithm.
  % similar to description in the work of Earl~\etal{}~\cite{shuyingliang:Earl:2012:IPDCFA}.
  
%  The next question is that how to work with pushdown reachability algorithm.
%  Since the root set for garbage collection depends on the entire stack,
%   it needs to access  (read) to the \emph{entire} stack during a transition.
%   This is in fact an \emph{introspective}  pushdown systems  that have read access
%   to the \emph{entire} stack during a transition, as described in the work of~\cite{shuyingliang:Earl:2012:IPDCFA}.
%  %
%   We presents the actual working algorithm in Alg.\ref{algo:stepIPDS} to show how to enable abstract garbage collection in the pushdown reachability algorithm, 
%      as to be illustrated in Section~\ref{subsec:fix-point-eval}.

%

\subsection{Abstract garbage collection enhanced with liveness analysis}\label{subsec:live}
Abstract garbage collection can avoid conflating abstract objects for reused variables or formal parameters,
but it can not discover ``garbage'' or ``dead'' abstract objects in the local scope.
The following example illustrates this:
\begin{center}
%\fontsize{9pt}{4pt}\selectfont
\begin{Verbatim}[fontsize=\relsize{-1}]
bool foo(A a) {
  B b = B.read(a);
  C p = C.doSomething(b); 
  return bar(C.not(p));
}
\end{Verbatim}
\end{center}
Obviously, in the function body \texttt{foo}, \texttt{b} is actually ``dead'' after the second line.
However, n\"{a}ive abstract garbage collection has no knowledge of this.
 In fact, this is a problem for n\"{a}ive concrete garbage collection~\cite{shuyingliang:Agesen:1998:GCL}.
 In the realm of static analysis, the garbage value pointed to by \texttt{b} can pollute the exploration of the entire state space.
 
 In addition, 
  in the register-based byte code that our implementation analyzes, 
 there are obvious cases  where the same register is reassigned multiple 
    times at different sites within a method. 
    The direct adaptation of abstract garbage collection to an object-oriented setting in Section~\ref{subsec:agcoo}
        cannot collect these registers between uses.
           %
            %For object-oriented programs, 
            %we want to collect registers that are unreachable that can be done by description in Section~\ref{subsec:agcoo},   but not without an intervening assignment.
            In other words, for object-oriented programs, we also want to collect ``dead'' registers, even though they are reachable under description in Section~\ref{subsec:agcoo}.
            This can be easily achieve by using liveness analysis. 
            %In fact, the additional benefits of  liveness leads us not to 
   % This problem can be easily solved by using liveness analysis.
Of course,   we could also solve it by transforming the byte code into Static Single Assignment (SSA) form. 
However, as mentioned above, liveness analysis has additional benefits, 
so we chose to enhance the abstract garbage collection with live variable analysis (LVA).

% 
%    As it turns out, the fix for this problem 
%    is a classic data-flow analysis:
%    live variable (register) analysis (LVA).
%      %
        LVA  computes the set of variables  that are \textit{alive} at each statement within a method.
The garbage collector can then more precisely collect each frame.

  Since LVA is well-defined in the literature~\cite{local:new-dragon}, 
  we skip the formalization here, 
  but   the $\mathit{Root}$  is now modified to collect
      only \textit{live}  variables of the current statement $\Lives(s_0)$:

      \vspace{-10pt}
      {\small
      \begin{align*} 
       \mathit{Root}(\vec{s}, \afp,\astore,\acont)  =
     \{(v',\afp) \} \union
    \mathit{StackRoot}(\acont), \\
    \text{ where }  (v', \afp) \in \mathit{dom}(\astore) ~\text{and}~ v' \in \Lives(s_0)
      \text.
      \end{align*} }%
 The liveness property is embedded in the overall \texttt{eagc} subroutine in Alg.~\ref{algo:stepIPDS}.

\section{Extending pushdown reachability analysis for exceptions}\label{sec:reacha}
Given the formalisms in the previous sections, 
it is not immediately clear how to convert these rules into a static analyzer,
or more importantly, how to handle the unbounded stack without it always visiting new machine configurations. 
Thus, we need a way to compute a finite summary of the reachable machine configurations.

In abstract interpretation frameworks, the Dyck State Graph synthesis
algorithm~\cite{shuyingliang:Earl:2010:Pushdown}, which is a purely functional
version of the Saturation algorithm~\cite{mattmight:Reps:1998:CFL},
 provides a method for computing reachable pushdown control states.
We build our algorithms on the work of Earl~\etal{}~\cite{shuyingliang:Earl:2012:IPDCFA}.
As it turns out, 
it is not hard to extend the summarization idea to deal with an unbounded stack with exceptions.
In the following sections, 
we present the complete algorithm in a top-down fashion, 
which aims to easily turn into actual working code. 
The algorithm code uses previous definitions specified in Section~\ref{sec:pushdown-sem}.

\subsection{Analysis setup}\label{subsec:setup}

\setlength{\intextsep}{0pt}
\setlength{\textfloatsep}{3pt}
\setlength{\floatsep}{2pt}
% top level
\begin{algorithm}[h]
{\fontsize{8pt}{6pt}\selectfont
\DontPrintSemicolon
\KwIn{$s$:\text{ a list of program statements (with an initial entry point $\stmt_0$).}}
\KwOut{\text{Dyck State Graph }$\DSG:$ a triple of a set of control states \text{ a set of 
edges, and a initial state.}}
$\astore_0 \longleftarrow$ \text{empty store}\;
$\fp_0 \longleftarrow $ \text{initial empty stack frame pointer}\;
$\atm_{0} \longleftarrow$ \text{empty list of contexts}\;
$\aqstate_{0} \longleftarrow (\stmt_0, \fp_0, \astore_0, \atm_0)$ \;
%$\aconf \longleftarrow (\aqstate, NIL)$\;
$ \text{The initial working set } \ToVisit_0 \longleftarrow \{\aqstate_0\}$\;
$\IECG_0  \longleftarrow (\emptyset,\emptyset,\emptyset,\emptyset,\emptyset,\emptyset )$\;
$\DSG_0 \longleftarrow (\{\aqstate_0\}, \emptyset, \aqstate_0)$\;
 ($\DSG, \IECG, \astore, \ToVisit$)$\longleftarrow$\text{\sc{Eval}}($\DSG_0, \IECG_0, \astore_0, \ToVisit_0$)\;
 \Return $\DSG$
 }
\caption{{\sc Analyze}}
\label{algo:analyze}
\end{algorithm}

The analysis for a program  starts from the \text{\sc{Analyze}} function, 
as shown in Alg.~\ref{algo:analyze}.
It accepts a program expression (an entry point to a program), 
and gives out a  Dyck State Graph (DSG).
Formally speaking, a DSG of a pushdown system is the subset of a pushdown system reachable over legal paths.
(A path is \emph{legal} if it never tries to pop $a$ when a frame other than $a$ is on top of the stack.)
%
%So our analysis first constructs 
%an initial DSG with the initial control state $\aqstate_0$ (Ln.1-5),
%an empty edge set (Ln.8) and
%a working set $\ToVisit_0$.
Note that the $\sa{Time}$ component is designed for accommodating traditional analysis, depending on actual implementation. 
For example, the last $k$ call sites  or object-allocation labels, or the mix of them.
The analysis produces DSG from the subroutine \text{\sc{Eval}}, 
which is the fix-point synthesis algorithm.

In Alg.~\ref{algo:analyze}, 
$\IECG$ is a composed data structure used in the $\epsilon$ summarization algorithm.
It is derived from the idea of an $\epsilon$ closure graph (ECG) in the work of Earl~\etal{}~\cite{shuyingliang:Earl:2010:Pushdown}, 
but supports efficient  caching of $\epsilon$ closures along with transitive push frames on the stack. 
Specifically, $\IECG$ = ($\EpsPreds, \EpsSuccs, \TopFrames, \PossibleStackFrames, \PredsForPush, \NonEpsPreds$). 
The six   components can be considered maps:
\begin{compactitem}
\item $\epsilon$ predecessors  $\EpsPreds$: $\aState \rightarrow \{\aState\}$, maps a target node to the source node(s) of an $\epsilon$ edge(s)
\item  $\epsilon$ successors $\EpsSuccs$: $\aState \rightarrow \{\aState\}$,
maps a source node to the target node(s) of an $\epsilon$ edge(s)
\item top frames $\TopFrames:  \aState \rightarrow \{\sa{Frame}\}$, records the shallow pushed stack frame(s) for a state node.
\item possible stack frames $\PossibleStackFrames: \aState \rightarrow \{\sa{Frame}\}$,  
compute  all   possible pushed stack frame of a state. It is used for abstract garbage collection.
\item predecessors for push action $\PredsForPush$: $(\aState, \sa{Frame}) \rightarrow \{\aState\}$, 
records  source state node(s) for a pushed frame and the net-changed state.
For example in the legal path: $\aqstate_0 \xrightarrow{g^+}  \aqstate_1 \longrightarrow ...   \xrightarrow{g^-}   \aqstate_2$, the entry $(\aqstate_1, g^+) \longrightarrow \{\aqstate_0\}  $  is in $\PredsForPush$.
\item non-$\epsilon$ predecessors ($\NonEpsPreds:$ $\aState \rightarrow \{\aState\}$), 
maps a state node to non-$\epsilon$ predecessors.
\end{compactitem}
These data structures (and $\IECG$)  have the same definition in the following algorithms.

 \subsubsection{Fix-point algorithm of the pushdown exception framework}\label{subsec:fix-point-eval}
Alg.~\ref{algo:eval} describes the fix-point computation for the reachability algorithm.
It iteratively constructs the reachable portion of the pushdown
transition relation (Ln.~5-12) 
by inserting $\epsilon$-summary edges whenever it finds empty-stack (Ln.~13-20) 
(\eg, push a, push b, pop b, pop a)
paths between control states.
\begin{algorithm}[h]
{\fontsize{8pt}{6pt}\selectfont
\DontPrintSemicolon
\KwIn{$\DSG, 
         \IECG ( \text{definition referred to Section~\ref{subsec:setup}}),
		% \EpsPreds, \EpsSuccs, 
		%	\TopFrames, \PossibleStackFrames, \PredsForPush, 
		%	\NonEpsPreds, 
			\astore, \text{working set }\ToVisit$}
\KwOut{$\DSG', 
         \IECG',
			\astore'', \ToVisit'$}
$\deltaStates, 
  \deltaEdges, 
  \astore'', 
  \ToVisit'
  \gets \emptyset $\;
 $(E, S, \aqstate_0) \gets \mathit{DSG} $\;
 $( \EpsPreds, \EpsSuccs, 
  \TopFrames, 
  \PossibleStackFrames, 
  \_, %\PredsForPush, 
 \_%	\NonEpsPreds
 	) \gets \mathit{IECG} $\;
 	$\IECG' \gets   (\emptyset,\emptyset,\emptyset,\emptyset,\emptyset,\emptyset )$\;
\For{$  s$ $\in$ $ \ToVisit $}{
  %% $\mathit{konts} \gets \TopFrames(s)$\;
  %% $\mathit{psf} \gets \PossibleStackFrames(s)$\;
  \For{$\akont$ $\in$ $\TopFrames(s)$}{ 
  		\For{$(g, s_1, \astore' ) \in \mathit{\text{\sc{Step}}(s, \akont, \PossibleStackFrames(s), \astore)}$} {
  			\If{  $ s \notin (\EpsSuccs(s) \cup S \cup \EpsPreds(s)) %\neg \text{inSeen}(s_1,  S, \EpsSuccs)  
  			%\land
  			    %  s \in \EpsPreds(s)%\neg\text{inEquivalenceClass}(s_1, \EpsPreds)
  			      $} {
  				\text{insert }$s_1$ \text{in} $\deltaStates$\;
  				\text{insert} $\mathit{E(s, g, s_1)}$ \text{in} $\deltaEdges$\;
  				\text{insert} $s_1$ \text{in} $\ToVisit'$\;
  				$\astore'' = \astore' \join \astore$
  			}
  		} 
  }
  \For{$E$ $\in$ $\deltaEdges$} {
  %	$\IECG'$ = \text{\sc{processEdge}}($E, \mathit{IECG}$)
   	\Switch{E}{
   		\Case {$(s', \varepsilon, s'')$}   {
   		$	\IECG' \gets $ \text{\sc Propagate}($E$, $\IECG$)
   		}
   		\Case{ $(s', g^+, s'')$} {
   			$	\IECG' \gets$	\text{\sc ProcessPush}($E, \IECG$) {
   			}
   		}
   	  \Case{ $(s', g^-, s'')$} {
   		   	$	\IECG' \gets$	 \text{\sc ProcessPop}($E, \IECG$) {
   	 }
   	 }
   	} 
  } 
$DSG' \gets (E \cup \deltaEdges, S \cup \deltaStates)$\;
  
   \If{$\astore''  \wk \astore \wedge \deltaEdges == \emptyset$}{
        \Return ($\mathit{DSG', \IECG', \astore'', \ToVisit'}$)
      }
      \Else{
        {\sc Eval}$(DSG', 
        				%\EpsPreds', 
          				% \EpsSuccs', \TopFrames', 
        				%  \PossibleStackFrames', 
        				%  \PredsForPush, 
        				 % \NonEpsPreds, 
        				 \IECG',
        				  \astore'', 
        				  \ToVisit')$\;
      }  
}
}
\caption{{\sc Eval}}
\label{algo:eval}
\end{algorithm}
Ln.~22-25 decides when to terminate the analysis:
no new frontier edges and the new store component $\astore''$ is subsumed by the old store $\astore'$.
The second condition uses the  technique presented by Shivers~\cite{shuyingliang:Shivers:1991:CFA}.
Otherwise, it recurs to $\text{\sc{Eval}}$.

% global widening
\begin{algorithm}[h]
{\fontsize{8pt}{6pt}\selectfont
\DontPrintSemicolon
\KwIn{\text{control state} $\aqstate$,  
\text{continuation} $\acont$, 
\text{a list of frames} $\vec{\aphrame}$}
\KwOut{\text{a set of records (\text{stack action }$g$, $\aqstate'$, $\astore$)}}
$\mathit{result} \longleftarrow \emptyset$\;
\For{(g, $\aqstate'$) $\in$ 	\text{\sc{StepIPDS}($\aqstate$, $\acont$, $\vec{\aphrame}$)}}{
	\text{insert} ($g$, $\aqstate'$, $\astore$) \text{in $\mathit{result}$}
}
\Return $\mathit{result}$
}
\caption{{\sc Step}}
\label{algo:step}
\end{algorithm}

\begin{algorithm}[h]
{\fontsize{8pt}{6pt}\selectfont
\DontPrintSemicolon
\KwIn{
$\text{a source state } \aqstate, \text{continuation }\acont, \text{list of frames }\vec{\aphrame} $,  
$\mathit{Options}$: global analysis options}
\KwOut{a set of tuples ($\aphrame', \aqstate'$)}
$\mathit{result}$ $\longleftarrow \emptyset$ \;
$\aqstate' \longleftarrow \aqstate$\;
\lIf{$\mathit{Options.doGC}$} {$\aqstate' \longleftarrow$ \texttt{eagc}($\aqstate, \aphrame$)}
$\mathit{confs} \longleftarrow$ \texttt{next}($\aqstate', \akont$)\;
\For{$(\aqstate'', \acont') \in \mathit{confs}$}{
	g $\longleftarrow$ \text{\sc{DecideStackAction}}($\acont, \acont'$)\;
	\text{insert} (\text{g}, $\aqstate''$) in $\mathit{result}$\;
}
\Return $\mathit{result}$
}
\caption{{\sc StepIPDS}}
\label{algo:stepIPDS}
\end{algorithm}

Now we explain Ln.~5-12 in more detail by  examining the subroutines that are called.
 As is shown in Ln.~7, 
 the \textit{raw} new states and edges are obtained  
from calling $\text{\sc{Step}}$ (shown in Alg.~\ref{algo:step}).
The algorithm enables the widening strategy in the pushdown reachability algorithm 
by instrumenting the $\astore$ component (it is \textit{widened} 
during iteration in \text{\sc{Eval}} (Ln.~7 and Ln.~12)).

The other important part of  the algorithm is \text{\sc{StepIPDS}},
Alg.~\ref{algo:stepIPDS} shows the details. 
\text{\sc{StepIPDS}}  does three things:
(1) It incorporates  the enhanced abstract garbage collection into the pushdown framework by calling \texttt{eagc} (Ln.~3).
The actual algorithm can be derived from 
the semantics 
presented in the Section~\ref{sec:eagc};
(2) It calls the pushdown abstract transition relation function of \texttt{next} based on the cleaned state after garbage collection. 
The semantics presented in Section~\ref{sec:pushdown-sem} 
 reflect the structure of \texttt{next};
(3) It summarizes the stack actions from the newly explored nodes,
 and so to construct possible edges for $\DSG$. 
 This is done in the algorithm \text{\sc{DecideStackAction}}, 
 which compares the continuation before the transition and the continuation after, 
 then decides which of the three stack actions: epsilon, push and pop to take.
Also note that we add only state nodes (e.g. $\aqstate$) into the working set 
%and to $\deltaStates$, 
if they are not appeared  in the following sets: 
state nodes of the current $\DSG$, 
 predecessors of $\aqstate$ and successors of $\aqstate$, 
 for the purpose of avoiding non-necessary re-computation.
% Pushdown system working with garbage collection
% Eval: the iteration
\begin{algorithm}[h]
{\fontsize{8pt}{6pt}\selectfont
\DontPrintSemicolon
\KwIn{\text{continuation before transition} $\acont$, \text{new continuation} $\acont'$}
\KwOut{\text{stack action} $g$}
\lIf{$\acont = \acont'$}{\Return  $\epsilon$}
($g_1$ \text{::} $\acont_1$) $\longleftarrow$  $\acont$ \;
($g_2$ \text{::} $\acont'_2$) $\longleftarrow \acont'$\;
\If{$\acont_1 == \acont'$}{
\Return  $g^-_1$}
\uElseIf{$\acont==\acont'_2$} {
\Return  $g^+_2$}
}
\caption{{\sc DecideStackAction}}
\label{algo:decideStackAction}
\end{algorithm}

Returning to \text{\sc{Eval}} in Alg.~\ref{algo:eval},
 Ln.~13-20 summarizes and propagates the new knowledge of the stack, given  $\deltaEdges$, by calling the Alg.~\ref{algo:equalize}, Alg.~\ref{algo:Processpop}, or Alg~\ref{algo:Processpush} based on the stack action.
 These algorithms are detailed in Section~\ref{subsec:dcg-exn}, along with the mechanism to deal with exceptions.

% %\input{algo_reachability}

\subsection{Synthesizing a Dyck State Graph with exceptional flow} \label{subsec:dcg-exn}

For pushdown analysis  \emph{without exception handling}, 
only two kinds of transitions
can cause a change to the set of $\epsilon$-$\mathit{predecessors}$ ($\EpsPreds$):
an intra-procedural empty-stack transition
and a frame-popping procedure return.
With the addition of $\mathbf{handle}$ frames to the stack, 
there are several new cases to consider 
for popping frames (and hence adding $\epsilon$-edges).

In the  following text, we first 
highlight how to
handle the exceptional flows during DSG synthesis, 
particularly as it relates
to maintaining $\epsilon$-summary edges.
Then we present the generalized  algorithms for these cases.
The figures in this section use a graphical scheme
for describing the cases for $\epsilon$-edge insertion.
Existing edges are solid lines, while the $\epsilon$-summary edges 
to be added are dotted lines.
%Stack alphabets $\exnframe, \callframe, \phrame$ are the concrete denotations of $\aexnframe, \acallframe, \aphrame$ (specified in Section~\ref{fig:java-abstract-state-space-pushdown}).
The superscripts of $+$ and $-$ on exception handler frame $\exnframe$, 
call frame $\callframe$ and general frame $\phrame$ mean push or pop actions of the correspondent frames.

\begin{figure}
\centering
\begin{subfigure}{0.45\linewidth}
\centering
\includegraphics[scale=0.8]{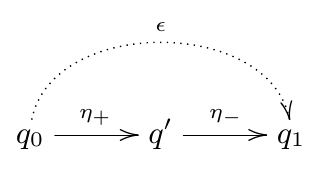}
\caption{Intra-procedural handler push/pop}
\label{fig:case1}
\end{subfigure}
\begin{subfigure}{0.45\linewidth}
\centering
\includegraphics[scale=0.8]{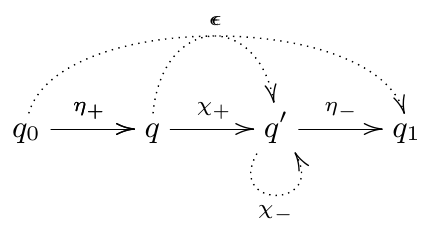}
\caption{Exception propagation}
\label{fig:case3}
\end{subfigure}
\begin{subfigure}{0.5\linewidth}
\centering
\includegraphics[scale=0.8]{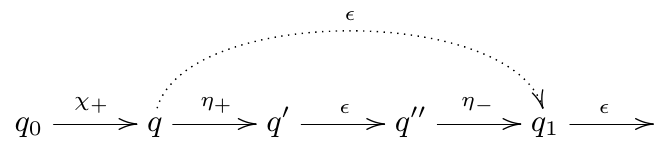}
\caption{Locally caught exceptions}
\label{fig:case2}
\end{subfigure}
\begin{subfigure}{0.5\linewidth}
\centering
\includegraphics[scale=0.8]{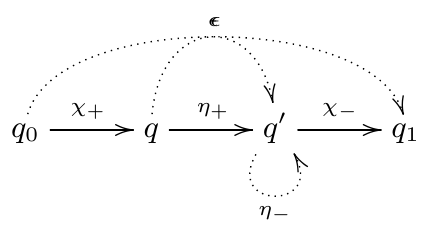}
\caption{Control transfers mixed in try/catch}
\label{fig:case4}
\end{subfigure}
\begin{subfigure}{0.4\linewidth}
\centering
\includegraphics[scale=0.8]{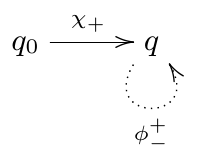}
\caption{Uncaught exceptions}
\label{fig:case5}
\end{subfigure}
\caption{Synthesizing a DSG with exceptional flow\\~~}
\label{fig:cases}
\end{figure}
\setlength{\textfloatsep}{3pt}
%\begin{compactitem}
%\item[Intraprocedural push/pop of handle frames]
\noindent\paragraph{Intraprocedural push/pop of handle frames}
The simplest case is entering a {\tt try} block 
and leaving a {\tt try} block entirely 
intraprocedurally---without throwing an exception.
Figure~\ref{fig:case1} shows such a case: if there is a handler push followed by a handler pop, 
the synthesized (dotted) edge must be added.\\
%
%\begin{figure} 
%\begin{equation}
%  \xymatrix{ 
%    \qstate_0 \ar[r]^{\exnframe_+} \ar @{.>} @(u,u) [rr]^\epsilon &
%    \qstate' 
%    \ar[r]^{\exnframe_-}
%   & \qstate_1
%  }  
%\end{equation}
%\caption{Intraprocedural handler push/pop}
%\label{fig:case1}
%\end{figure}
%\item[Locally caught exceptions]
\paragraph{Locally caught exceptions}
Figure~\ref{fig:case2} presents a case where 
a local handler catches an exception, popping it off the stack
and continuing.\\
%
%
%\begin{figure} 
%\begin{equation*}
%  \xymatrix{ 
%    \qstate_0 \ar[r]^{\callframe_+} &
%    \qstate \ar[r]^{\exnframe_+}  \ar @{.>} @(u,u) [rrr]^\epsilon&
%    \qstate' \ar[r]^{\epsilon} &
%   \qstate'' \ar[r]^{\exnframe_-}
%   & \qstate_1
%     \ar[r]^{\epsilon} &
%  }  
%\end{equation*}
%\caption{Locally caught exceptions}
%\label{fig:case2}
%\end{figure}
%\item[Exception propagation along the stack]
\paragraph{Exception propagation along the stack}
Figure~\ref{fig:case3} illustrates a case where 
an exception is not handled locally, and must pop off a 
call frame to reach the next handler on the stack.
In this case,
a popping self-edge from control state $q'$ to $q'$ 
lets the control state $q'$ see frames beneath the top.
Using popping self-edges, a single state can pop off as many frames as necessary 
to reach the handle---one at a time.\\
%\begin{figure} 
%\begin{equation*}
%  \xymatrix{ 
%    \qstate_0 \ar[r]^{\exnframe_+}\ar[r]^{\exnframe_+} \ar @{.>} @(u,u) [rrr]^\epsilon &
%    \qstate \ar[r]^{\callframe_+}  \ar @{.>} @(u,u) [r]^\epsilon&
%    \qstate' \ar[r]^{\exnframe_-} \ar @{.>} @(dl,dr)_{\callframe_-} &
%   \qstate_1
%  } 
%\end{equation*}
%\caption{Exception propagation}
%\label{fig:case3}
%\end{figure}%
%\item[Control transfers mixed in try/catch]
\paragraph{Control transfers mixed in try/catch}
Figure~\ref{fig:case4} illustrates the situation where a procedure tries to
return while a $\mathbf{handle}$ frame is on the top of the stack.
%
%\begin{figure} 
%\begin{equation*}
%  \xymatrix{ 
%    \qstate_0 \ar[r]^{\callframe_+}  \ar @{.>} @(u,u) [rrr]^\epsilon &
%    \qstate \ar[r]^{\exnframe_+}  \ar @{.>} @(u,u) [r]^\epsilon&
%    \qstate' \ar @{.>} @(dl,dr)_{\exnframe_-} \ar[r]^{\callframe_-}&
%    \qstate_1
%  } 
%\end{equation*}
%\caption{Control transfers mixed in try/catch}
%\label{fig:case4}
%\end{figure}
It uses popping self-edges as well to find the top-most $\mathbf{call}$ frame.\\
%\item[Uncaught exceptions]
\paragraph{Uncaught exceptions}
The case in Figure~\ref{fig:case5} 
shows popping all frames back to  
the bottom of the stack---indicating an uncaught exception.

\subsection{The generalized algorithms: \text{\sc{Propagate}}, \text{\sc{ProcessPop}}, \text{\sc{ProcessPush}}}\label{subsec:generalize}

Section~\ref{subsec:dcg-exn} graphically illustrates the new cases for handling exceptions (Figure~\ref{fig:cases}).
The following text presents the working algorithms to achieve the synthesis process.
Alg.~\ref{algo:equalize} handles the cases
 when an $\epsilon$ edge is added.
 These cases are: 
intra-procedural empty-stack transition, 
 a frame-popping procedure return, or
 a frame-popping intra-procedural or inter-procedural exception catch, 
 as presented in Figure~\ref{fig:cases}.
  \begin{algorithm}
 {\fontsize{8pt}{6pt}\selectfont
 \DontPrintSemicolon
  \KwIn{\text{An edge}$E$, an $\IECG$ (refer to Section~\ref{subsec:setup})}
  \KwOut{$\IECG'$}
   $( \EpsPreds, \EpsSuccs, 
   \TopFrames, 
   \PossibleStackFrames, 
    \PredsForPush, 
  	\NonEpsPreds
   	) \gets \mathit{IECG} $\;
   	$\mathit{topFramesToAdd} \gets \emptyset$\;
   	$\EpsPreds', \EpsSuccs', 
   	 \TopFrames', 
   	 \PossibleStackFrames', 
   	  \PredsForPush', 
   		\NonEpsPreds \gets \emptyset$\;
   		$(s_1, \varepsilon, s_2) \gets E$\;
  %   	\Switch{E}{
     	%	\Case {$(s', \varepsilon, s'')$}   {
     			$\mathit{preds} \gets \EpsPreds(s_1) \cup \{s_1\}$\;
     			$\mathit{nexts} \gets \EpsSuccs(s_2) \cup\{s_2\}$\;
     			\For{$s$ $\in$ $\mathit{preds}$}{ 	
     				$\EpsSuccs' \gets \EpsSuccs \sqcup [s \mapsto \EpsSuccs(s) \cup \mathit{nexts} ]$\;
     				%\text{insert} ($s, \EpsSuccs(s) \cup \mathit{nexts}$) \text{in} $\EpsSuccs$\;
     				\text{insert} $\TopFrames(s)$ \text{in} $	\mathit{topFramesToAdd}$
     			}
     			\For {$s$  $\in$ $\mathit{nexts}$}{
     				$\EpsPreds' \gets \EpsPreds \sqcup [s \mapsto \EpsPreds(s) \cup \mathit{preds} ]$\;
     			%	\text{insert} ($s, \EpsPreds(s) \cup \mathit{preds}$) \text{in} $\EpsPreds$\;
     				$\TopFrames' \gets \TopFrames \sqcup [s \mapsto \TopFrames(s) \cup \mathit{topFramesToAdd} ]$\;
     				%	\text{insert} ($s, \TopFrames(s) \cup \mathit{topFramesToAdd}$) \text{in} $\ \TopFrames$\; 
     				\For{$f \in \TopFrames'(s_1)$}{
     					$\PredsForPush' \gets \PredsForPush \sqcup 
     					[(s, f) \mapsto \PredsForPush(s,f)]$\;
     				}
     				$\PossibleStackFrames' \gets ${{\sc updatePSF}$(s, \TopFrames', \PossibleStackFrames, 
 \NonEpsPreds, \EpsPreds')$}\;
     			}
  $\IECG' \gets (\EpsPreds', \EpsSuccs', \TopFrames', \PossibleStackFrames', \PredsForPush', \NonEpsPreds)$\;
  \Return $\IECG'$
  \SetKwProg{myproc}{Function}{}{}
  %  \myproc{propagate}{
  %  \KwIn{s, $\EpsPreds$}
  %  \KwOut{Boolean}
  %  \Return s $\in \EpsPreds(s)$\; 
  %  }
  }
  \caption{{\sc Propagate}}
  \label{algo:equalize}
  \end{algorithm}
 
 The algorithm works as follows:
 It accepts an $\epsilon$ edge E and the current record of $\IECG$ (introduced in Section~\ref{subsec:setup}) and produces a new IECG $\IECG'$.
 It propagates the $\epsilon$ successors for each control state in $\EpsPreds(s_1)\cup s_1$ (Ln.~8) and  prepares the accumulated  top frames for propagation for each successor state node in $\EpsSuccs$ (Ln.~12). 
 Similarly, it propagates the $\epsilon$ predecessors for each control state in  $\EpsSuccs(s_2)\cup s_2$. 
 The predecessor nodes of pushed frames for the current target note state $s$
will also be propagated with the new propagated top frames (Ln.~13-14).
 Finally, it propagates the possible stack frames $\PossibleStackFrames$ (for abstract garbage collection)  in Ln.~15,
  for each control state in the  original non-$\epsilon$ predecessors and new $\epsilon$ predecessors $\EpsPreds$, as shown in Alg.~\ref{algo:UpdatePSF} Ln.~2-3.

 \begin{algorithm}
 {\fontsize{8pt}{6pt}\selectfont
  \DontPrintSemicolon % Some LaTeX compilers require you to use \dontprintsemicolon instead
  \KwIn{ $s, \TopFrames', 
  \PossibleStackFrames, 
  \NonEpsPreds, 
  \EpsPreds'$}
  \KwOut{$\PossibleStackFrames''$ }
  $\PossibleStackFrames' \gets 
  \PossibleStackFrames \sqcup [s \mapsto \TopFrames'(s)]$\;
  \For {$\mathit{spred} \in \NonEpsPreds(s) \cup \EpsPreds'(s)$} {
  	$\PossibleStackFrames'' \gets \PossibleStackFrames' \sqcup [s \mapsto \PossibleStackFrames'(\mathit{spred})]$
  }
   \Return $\PossibleStackFrames''$
   }
  \caption{{\sc UpdatePSF}}
  \label{algo:UpdatePSF}
  \end{algorithm}

Alg.~\ref{algo:Processpop} handles the case of popping frames, including function call return popping and exception handling popping.
The algorithm  is reduced to  Alg.~\ref{algo:equalize} to introduce $\epsilon$ edges, 
for each tuple in $\PredsForPush$.

 \begin{algorithm}
 {\fontsize{8pt}{6pt}\selectfont
 \DontPrintSemicolon % Some LaTeX compilers require you to use \dontprintsemicolon instead
 \KwIn{$E, \IECG$}
 \KwOut{$\IECG'$}
 $\IECG' \gets \emptyset$
  		$(s_1, g^-, s_2) \gets E$\;
 %$\PossibleStackFrames' \gets 
 %\PossibleStackFrames \sqcup [s \mapsto \TopFrames'(s)]$\;
 \For {$\mathit{s} \in \PredsForPush(s_1, g^-)   $} {
  	$\IECG' \gets $ $\IECG$ $\sqcup$ {\sc Propagate}($(s, \varepsilon, s_2), \IECG)$
 }
 
 \Return $\IECG'$
  }
 \caption{{\sc Processpop}}
 \label{algo:Processpop}
 \end{algorithm}
 
Alg.~\ref{algo:Processpush} is presented for completeness. 
 \begin{algorithm}
 {\fontsize{8pt}{6pt}\selectfont
 \DontPrintSemicolon % Some LaTeX compilers require you to use \dontprintsemicolon instead
 \KwIn{$E, \IECG$}
 \KwOut{$\IECG'$}
 {$\IECG'$}
  $( \EpsPreds, \EpsSuccs, 
  \TopFrames, 
  \PossibleStackFrames, 
   %\PredsForPush, 
 	\NonEpsPreds
  	) \gets \mathit{IECG} $\; 
  	$\EpsPreds', \EpsSuccs', 
  	 \TopFrames', 
  	 \PossibleStackFrames', 
  	  \PredsForPush', 
  		\NonEpsPreds' \gets \emptyset$\;
  		$(s_1, g^+, s_2) \gets E$\;
 %$\PossibleStackFrames' \gets 
 %\PossibleStackFrames \sqcup [s \mapsto \TopFrames'(s)]$\;
 \For {$\mathit{s}\in\EpsSuccs(s_2) \cup \{s2\}  $} {
 	$\TopFrames' \gets \TopFrames \sqcup [s \mapsto \{f\}]$ \;
 	$\PredsForPush' \gets \PredsForPush \sqcup [(s,f) \mapsto \{s_1\}]$\;
  $\NonEpsPreds' \gets \NonEpsPreds \sqcup [s \mapsto \{s_1\}]$\;
 	$\PossibleStackFrames' \gets ${{\sc UpdatePSF}$(s, \TopFrames', \PossibleStackFrames, \NonEpsPreds', \EpsPreds)$}\;
 }
 $\IECG' \gets (\EpsPreds, \EpsSuccs, \TopFrames', \PossibleStackFrames', \PredsForPush', \NonEpsPreds')$\;
 \Return $\IECG'$
  }
 \caption{{\sc Processpush}}
 \label{algo:Processpush}
 \end{algorithm}
It handles pushing stack frames in function calls and exception handlers in $\texttt{try}$ blocks.
Since the pushing action introduces a new top frame, 
it  maintains extensions (propagation) for the data structure top frames $\TopFrames$, 
predecessors for push frames $\PredsForPush$, 
non-$\epsilon$ predecessors $\NonEpsPreds$ 
and possible stack frames $\PossibleStackFrames$.

\section{Implementation}\label{sec:implemenation}
% bytecode. 
% finally.
 We have implemented the analysis framework\footnote{\url{https://github.com/shuyingliang/pushdownoo}}
with pushdown abstraction and enhanced abstract garbage collection to analyze Android applications, which are Java programs.
The analyzer works directly on Dalvik bytecode, 
which is compiled from Java programs into Dalvik Virtual Machine (DVM).
Different from Java bytecode, Dalvik bytecode is register-based. 
What's more important, 
it closely resembles the high-level Java source code. 
We choose to work on bytecode in implementation for two reasons:
(1) The semantics of Dalvik bytecode is 
almost identical to  that of high-level Java 
while bringing  more advantage in analyzing \texttt{finally}.
(2) It enables us to analyze off-the-shelf Android 
applications.

%(2) It simplifies the analysis of \texttt{finally},
% as illustrated in the following section; 

\textbf{The \texttt{finally} blocks:}
In previous sections, 
we described the semantics and algorithms for \texttt{try/catch}.
To analyze full-featured exceptions, we have to deal with the \texttt{finally} blocks. It is known to be non-trivial 
to handle \texttt{finally} in static analysis~\cite{shuyingliang:Chatterjee:2001:CPA}. 
However, this is not a problem in our analysis.
The reason is that the analyzer directly works on object-oriented byte code, 
where the \texttt{finally} is compiled away by compiler in this level.
Specifically, 
the blocks of code for \texttt{finally} are
copied into  try and catch blocks before any possible exit points, 
which include normal \texttt{return} statements or \texttt{throw} statements.
This eases the static analysis substantially.
In addition,  \texttt{finally} blocks are translated as one kind of \texttt{catch} handler, which is the \texttt{catchall} handler, with the exception type \texttt{java/lang/Exception}. 
During the pushdown analysis, \texttt{catchall} is placed below any other normal \texttt{catch} handlers on the stack,  it is matched last and executed for any possible \texttt{throw} exceptions.

\section{Evaluation}\label{sec:evaluation}

%To evaluate the effectiveness of our analysis technique for exception-flow analysis,
%we compare our analysis with one of the well-known finite-state based static
%analysis frameworks---WALA.\footnote{\url{http://sourceforge.net/projects/wala/}}.

To evaluate the effectiveness of our analysis technique,
we compare our analysis with one of the well-known finite-state based static
analysis frameworks---WALA.\footnote{\url{http://sourceforge.net/projects/wala/}}
%we made our efforts to compare with some state of art static analysis frameworks and make the comparison faithfully.
In fact, there are two representative traditional static analysis frameworks for object-oriented programs, 
Doop~\cite{Bravebboer:2009:declare-pointsto} and WALA. %\footnote{\url{http://sourceforge.net/projects/wala/}}
They are both finite-state static analysis but orthogonal work. 
For this reason, there are no comparison results reported in the literature for the two analysis frameworks.
However, we still experimented with  Doop~\cite{Bravebboer:2009:declare-pointsto}  virtual image provided by the Doop authors.
However, the results were incomplete due to significantly slower running times
on several of the DaCapo~\cite{local:DaCapo:paper} benchmarks.  As a result, we do not feel a fair
comparison can be made.

As it turns out, WALA is  based on the work of Reps~\etal{}~\cite{mattmight:Reps:1998:CFL}, 
which was later formalized into pushdown reachability.
In this sense, WALA is more similar to our approach with respect to pushdown reachability.
Therefore, we compare our analysis with WALA. 
In specific, WALA  mainly adopts co-analysis of control-flow and data-flow analyses, 
performing  call-graph construction and pointer analysis together,
by propagating pointer information on the constructed CFG. 
The framework provides several context-sensitivities~\cite{local:alias-WALA}, 
including 
%Rapid Type Analysis (RTA), 
0-CFA, 0-1-CFA (0-CFA with 1-object sensitivity),
%Vanilla-0-1-CFA (an unoptimized version of 0-1-CFA), 
and analysis with additional
disambiguation of container elements 0-container and 0-1-container. \footnote{http://wala.sourceforge.net/wiki/index.php/UserGuide:PointerAnalysis}
In particular, the 0-1-CFA enables several optimizations for string and thrown objects. %\footnote{http://wala.sourceforge.net/wiki/index.php/UserGuide:PointerAnalysis}, \etc{}.
  The 0-1-container policy extends the 0-1-CFA 
   with unlimited object-sensitivity for collection objects, 
   which is the most precise default option.
Our evaluation uses the 0-1-container as the baseline.

To make the comparison more compelling, 
we conduct experiments on the DaCapo~\cite{local:DaCapo:paper} benchmarks.
It has much larger scale code bases to analyze than ordinary Java applications presented in the Google market. 
This allows a more realistic workload to stress-test the analysis.
Due to some conflicts in Java GUI classes, \texttt{eclipse} can not be ported in DVM.
Other 10 programs out of 11 Java applications in the DaCapo benchmark suffice for our purpose.

%Since the analyzer works directly on Dalvik bytecode, 
%which is compiled from Java programs in Dalvik Virtual Machine (DVM), 
%we have successfully compiled 10 out of 11 Java applications in the DaCapo benchmark (v.2006-10.MR2)  in DVM
%with minor changes in the source code (mainly \texttt{enum} is changed to another name because \texttt{enum} is a keyword in JRE 1.5 or later).
%We encapsulate the \texttt{main} method of each benchmark in the entry point \texttt{onCreate} of a class of type \texttt{Activity}.
%These benchmarks are compiled using the built-in tool \texttt{dx} in the Android SDK.
% Some GUI class references (especially \texttt{awt}) in Java programs are resolved by including \texttt{rt.jar} in the Android class path. 
% To avoid name space conflicts in packages and classes, we use
% \texttt{jarjar}\footnote{\url{https://code.google.com/p/jarjar}}
% to repackage some Java standard libraries that are re-implemented in Android.
%The only Java program that is not ported is \texttt{eclipse}, 
%which involves substantial conflicts in Java GUI classes (\texttt{awt, swing, swt}). 
%We believe the other ten programs suffice for our purpose.

%
\subsection{Metrics for precision}
Our basis for comparison in precision
 is the average cardinality of a points-to set~\cite{Fu:2005:rubust-java-server-apps,Bravenboer:2009:Exceptions,shuyingliang:2013:Hybrid-points-to} and exception-catcher links (E-C links)~\cite{Fu:2005:rubust-java-server-apps}.
%  to reflect the precision of handling exceptional flows. 
 
The average cardinality of a points-to set computes the average number of abstract (exception) objects for pointers 
that are collected into a single representative in the abstraction.
In our evaluation, it has two forms: \texttt{VarPointsTo} and \texttt{Throws}.
\texttt{VarPointsTo} refers to the average cardinality of the points-to set for non-exception abstract objects,
and \texttt{Throws} refers to exception objects specifically. 
(In Table~\ref{tbl:precision-result}, 
we normalized the two metrics computed in WALA, relative to that in our analysis.).
We adopt this metric because it reflects analysis precision by recognizing that
the more objects are conflated for a variable, the less precise the analysis.
When this metric is a large value, it indicates a negative impact on
normal control-flow analysis 
because it means that virtual method resolution needs to dynamically dispatch to  more 
than one function causing spurious control-flow paths. 
This same reasoning applies for exception-flow analysis.
(The more subtle relationships have been illustrated in Section~\ref{sec:intro}).
More rationals of using  this metric to measure precision for object-oriented programs are illustrated in the  literature~\cite{Fu:2005:rubust-java-server-apps,Bravenboer:2009:Exceptions,shuyingliang:2013:Hybrid-points-to}.
Following WALA's heap model,
%\footnote{%
%In WALA, the pointer-to relation is computed 
%from {\tt PointerKey} to a set of {\tt InstanceKey}s, 
%where a {\tt PointerKey} may represent a local variable, a static field, or an instance field of objects from a particular allocation site, 
%and an {\tt InstanceKey} may represent all objects of a particular type, all objects from a particular allocation site, all objects from a particular allocation site in a particular context, or other variants.},
we compute the same metric in our pushdown framework.

%In addition to the \texttt{Throws} metric,
The E-C links, proposed by Fu~\etal{}~\cite{Fu:2005:rubust-java-server-apps} is to reflect the 
precision of handling exceptional flows. It is also used in the work of~\cite{Bravenboer:2009:Exceptions}.
We compute the metric in our analysis framework, which is within the
range of 1-3 across the DaCapo benchmarks. 
Because WALA directly computes the catchers intra-procedurally,
we do not compute the comparison ratio  as we do for \texttt{VarPointsTo} and \texttt{Throws}.

In addition, we also evaluated  the precision of our pruned, pushdown analysis with respect to the client security analysis. We refer readers to the related work~\cite{shuyingliang:Liang:2013:pushdownandroid}.

\input{benchmarks-table}

\subsection{Results}
Table~\ref{tbl:precision-result} shows that
the pushdown exception-flow analysis with enhanced abstract garbage \texttt{pdxfa+eagc}
outperforms  finite-state context-sensitive analysis (represented by WALA) 
with a precision of 4.5-11 times for \texttt{Throws} and up to 7 times for general points-to information \texttt{VarPointsTo}.
\texttt{Nodes} and  \texttt{Edges} are control-flow graph information.
 \texttt{Methods} denotes the analyzed methods.
 The values in these columns in Table~\ref{tbl:precision-result} 
 are normalized relative to those reported by WALA 0-1-contain analysis.
 As is shown in Table~\ref{tbl:precision-result}, 
 our pruned, pushdown analysis technique (\texttt{pdxfa+eagc}) 
 generally explores more edges and nodes, 
 and explores up to 3.4 times more methods.
 %\footnote{The major reason that the number of methods discovered is different from what WALA discovers is that  WALA filters out some common methods (e.g. I/O API), as specified in "AnalysisScope".}

To evaluate the contribution of each aspect 
(pushdown exception-flow analysis and enhanced abstract garbage collection) to precision improvement, when
comparing with WALA, 
we also conduct an additional experiment with only the pushdown exception-flow analysis with 1-object sensitivity (as WALA 0-1-container does), denoted as the option \texttt{pdxfa+1obj}.
The result shows that the \texttt{pdxfa} improves the precision more than enhanced abstract garbage collection does.

\input{benchmarks-table-time}
\subsection{Analysis time}

For completeness, we also report an analysis time comparison. 
Table~\ref{tbl:result-time}  is the ratio of our analysis time to that of WALA.

WALA reports less analysis time than our analysis.
This is not surprising. 
First, our analysis is derived from the polynomial complexity algorithm in~\cite{mattmight:Reps:1998:CFL,shuyingliang:Earl:2012:IPDCFA}.
Even with enhanced garbage collection, it only reduces the complexity by a constant factor. 
Second, WALA has been significantly optimized by the IBM research lab,
particularly with underlying Java (collection) libraries   rewritten   specifically for  its framework.
Our implementation is based on Scala's default data structures 
and our specialized G\"odel hashing data structure~\cite{shuyingliang:godel}.
%It is not as optimized as WALA.
Last but not least, the analysis time is reasonably acceptable,  
given the high precision that our analysis technique can provide.
For example, for the largest benchmark \texttt{chart}, 
the unoptimized analyzer takes roughly 13 minutes.

\section{Related work}
\label{sec:related}

%% \paragraph{\textit{Exception Analysis}}
\textbf{\textit{Exception Analysis}}~~
The bulk of the earlier literature for analyzing Java programs
has generally focused on finite-state abstractions, \ie,
 $k$-CFA and its variants. 
Specifically, for the work that acknowledges exceptional flows, the analysis 
is based on either context-insensitivity or a limited form of context-sensitivity.
Analyzers that use only syntactic, type-based analysis of exceptional flow
are extremely imprecise~\cite{Leroy:2004:TypeBasedUncaughtExceptions,Robillard:2003:evolution-exception}.
Propagating exceptions via the imprecise call graphs cause the analysis result in:
(1)  inclusion of many spurious paths between exception throw sites and handlers 
that are not truly realizable at run time;
(2)  unable to tell and differentiate where an exception comes from. 
% 
%WALA\cite{local:wala:url} uses only declared types to determine exceptional flow, 
%and assuming every Potentially Excepting Instruction (PEI) might throw exceptions. 
%It essentially has no sufficient contexts to remove infeasible exceptional control flow.
Fu~\etal{}~\cite{Fu:2005:rubust-java-server-apps}  approached the problem 
   by employing  points-to information to refine
   control-flow reachability. 
Later,
Bravenboer and Smaragdakis exploited 
this mutual recursion by
co-analyzing data- and exception-flow~\cite{Bravenboer:2009:Exceptions}.
It reports precision improvement in both pointer-to analysis and exception analysis.
 
%Prabhu~\etal{}~\cite{Prabhu:2011:IEA} have proposed a modular refinement to construct control graph
%to analyze interprocedual exceptions for C++ programs.  
%They avoid points-to analysis,
%opting to use
%static type information to decide which catch handlers is invoked.
%But their result can be further improved by applying pointer analysis to functional pointers and virtual function calls.
% In addition, their stack unwinding issue could be resolved within pushdown scheme.
%One interesting thing worth pointing out is that  the paper mentions that stack unwinding is an major issue in C++, 
%because when an exception escapes out of a function, 
%destructor are invoked on all stack allocated objects 
%between the occurrence of the exception and the catch handler in a process~\cite{Prabhu:2011:IEA}. 
% Intuitively, this issue can be resolved naturally
%  within the pushdown scheme with the stack faithful models the control flows and exceptions flows.

%% \paragraph{\textit{Points-to Analysis}}
\textbf{\textit{Points-to Analysis}}~~
Precise and scalable context-sensitive points-to analysis 
has been an open problem for decades.
We describe a portion of the representative work in the literature.
%
%Progress in general has been gradual,
%with results like object-sensitivity~\cite{local:Milanova:2007:LCP,local:Milanova:2005:parameterizedobject}
% intermittently providing a leap for most programs.
%%
%Most results target improvements for individual classes of programs.
%%
%Our analyses targets at all programs, 
%and it is orthogonal to and compatible with
%results like object-sensitivity.
%
Much work in pointer analysis exploits methods to improve performance
by strategically reducing precision. 
Lattner~\etal~show that an analysis with a context-sensitive
heap abstraction can be efficient by sacrificing precision 
under unification constraints~\cite{local:Lattner:2007:MCP}. 
In full-context-sensitive pointer analysis, 
%the literature has sought
% context abstractions that provide precise pointer information
%while not sacrificing performance.
Milanova~\etal{}~found that an object-sensitive analysis~\cite{local:Milanova:2005:parameterizedobject}
 is an effective context abstraction for object-oriented programs. 
 %% This is confirmed by the extensive evaluation by Lhot\'{a}k~\cite{local:Lhotak:2008:EBC}.
BDDs have been used to compactly represent the large  amount of redundant  data in context-sensitive pointer analysis 
efficiently~\cite{local:Berndl:2003:PAU,Whaley:2004:CCP,local:Xu:2008:MEC}. 
%Specifically, Xu and Routev's work~\cite{local:Xu:2008:MEC} reduces the redundancy by choosing the right context abstractions.
%
Such advancements could be applied to our pushdown framework, as they are orthogonal to its central thesis.
Recently, Khedker~\etal~\cite{Khedker:2012:LPA} exploits  liveness analyses to improve points-to analysis.
Our work also uses liveness analyses but extends it to work with abstract garbage collection.
In fact, to the best of our knowledge,
we are the first work that explores abstract garbage collection in analyzing object-oriented programs and enhances it with liveness analysis to explicitly prune points-to precision.

%% \paragraph{\textit{Pushdown analysis for the $\lambda$-calculus}}
 \textbf{\textit{Pushdown analysis for the $\lambda$-calculus}}~~
Vardoulakis and Shivers's CFA2~\cite{shuingliang:vardoulakis-diss12}
is the precursor to the pushdown control-flow
analysis~\cite{shuyingliang:Earl:2010:Pushdown}.
%
%CFA2 is a table-driven summarization algorithm that exploits the
%balanced nature of calls and returns to improve return-flow precision
%in a control-flow analysis.
%%
%While CFA2 uses a concept called ``summarization,'' it is a
%summarization of execution paths of the analysis, roughly equivalent
%to Dyck state graphs.
%
%In terms of recovering precision, pushdown control-flow
%analysis~\cite{shuyingliang:Earl:2010:Pushdown} is the dual to abstract garbage
%collection:
%%
%it focuses on the global interactions of configurations via
%transitions to precisely match push-pop/call-return, thereby
%eliminating all return-flow merging.
%%
%However, pushdown control-flow analysis does nothing to improve
%argument merging.
Our work directly draws on the work 
of  pushdown analysis for higher-order programs~\cite{shuyingliang:Earl:2010:Pushdown} and
 introspective pushdown system (IPDS) for higher-order programs~\cite{shuyingliang:Earl:2012:IPDCFA}.
We extend the earlier work in three dimensions: 
(1)  We generalize the framework to an object-oriented language;
 (2) We adapt the Dyck state graph synthesis algorithm to handle the new stack change behavior
introduced by exceptions;
 (3) We reveal necessary details to design and implement a static analyzer even in the exceptions.

\vspace{-.3em}

%% \paragraph{\textit{CFL- and pushdown-reachability techniques}}
\textbf{\textit{CFL- and pushdown-reachability techniques}}~~
Earl~\etal~\cite{shuyingliang:Earl:2012:IPDCFA}
develop a pushdown reachability algorithm
suitable for the pushdown systems that we generate.
It essentially draws on CFL- and pushdown-reachability
analysis~\cite{mattmight:Bouajjani:1997:PDA-Reachability,mattmight:Kodumal:2004:CFL,mattmight:Reps:1998:CFL,mattmight:Reps:2005:Weighted-PDA}.
For instance, epsilon closure graphs,
or equivalent variants thereof, appear in many
context-free-language and pushdown reachability algorithms.
Dyck state graph synthesis is an attractive perspective
on pushdown reachability because it allows targeted modifications to the algorithm.

%CFL-reachability techniques have also been used to compute classical
%finite-state abstraction CFAs~\cite{mattmight:Melski:2000:CFL} and
%type-based polymorphic control-flow
%analysis~\cite{mattmight:Rehof:2001:TypeBased}.
%%
%These analyses should not be confused with pushdown control-flow
%analysis, which is computing a fundamentally different kind of CFA.

%% \paragraph{\textit{Pushdown exception-flow analysis}}
\textbf{\textit{Pushdown exception-flow analysis}}~~
There is few work on pushdown analysis for object-oriented languages as a whole.
Sridharan and Bodik proposed demand-driven analysis for Java that
matches reads with writes to object fields selectively, by using
refinement~\cite{Manu:2006:RefinementJava}.  They employ a
refinement-based CFL-reachability technique that refines calls and
returns to valid matching pairs, but approximates for recursive calls.
They do not consider specific applications of CFL-reachability to
exception-flow.

\section{Conclusion}\label{sec:conclusion}
%

%Statically reasoning in the presence of exceptions and about the effects of
%exceptions is challenging: 
%%
%exception-flows are mutually determined by traditional control-flow analysis
%and points-to analysis. 
%%
%We tackle the challenge of analyzing exception-flows from two angles.
%First, from the angle of pruning precise control-flows (both normal and exceptional),
%we derive a pushdown analytic framework to object-oriented programs 
%in the presence of full-featured exceptions.
%Pushdown analyses model the program stack using the unbounded stack of a
%pushdown system, which, unlike traditional analyses, 
%allows them to precisely match throwers to catchers.
%%
%Second, from the angle of pruning precise points-to information,
%we generalize \textit{abstract} garbage collection 
%to object-oriented programs and 
%enhance  it with live variable analysis.
%%
%We then seamlessly weave the techniques together, 
%yielding highly precise exception-flow analysis,
%without becoming intractable, even for large applications.
% We evaluate our pruned, pushdown exception-flow analysis, 
% comparing it with an established, 
% traditional analysis, on large scale standard Java benchmarks.
% The results show that our pruned, pushdown exception-flow analysis 
% \textit{significantly} improves  analysis precision over  traditional analysis
% within a reasonable analysis time.
% 
%%for now
%Poor analysis of exceptions pollutes the inter-procedural control-flow analysis
%of a program.
%
Exception-flows are mutually determined by traditional control-flow analysis
and points-to analysis. 
In order to model exceptional control-flow precisely, 
we abandoned traditional
finite-state approaches (e.g. $k$-CFA and its variants).
In its place, 
we generalized pushdown control-flow analysis from the
$\lambda$-calculus~\cite{shuyingliang:Earl:2012:IPDCFA} to object-oriented programs, 
and made it capable of handling exceptions in the process.
Pushdown control-flow analysis models the program stack (precisely) with the
pushdown stack, for the purpose of pruning control-flows.
To prune the precision with respect to  points-to information, 
we  adapted abstract garbage collection to object-oriented
program analysis and enhanced it with live variable analysis.
Computing the  reachable control states of the pushdown system 
(the enhanced Dyck state graph) yields 
combined data-flow analysis and control-flow analysis of a program.
Comparing this approach to the established traditional analysis framework shows
\textit{substantially} improved precision, within a reasonable analysis time.

\section{Acknowledgments}
This material is based on research sponsored by DARPA under agreement
number FA8750-12-2-0106. The U.S. Government is authorized to
reproduce and distribute reprints for Governmental purposes
notwithstanding any copyright notation thereon.

%% file: benchmarks-table.tex
% % % current results, I will consider to add time. too.
\begin{table*}
\centering
{\footnotesize
    \begin{tabular}{ c| c | c| c | c | c | c | c }
    \hline
    \textbf{Benchmark} & \textbf{LOC} & \textbf{Opts} & \textbf{Nodes} & \textbf{Edges} & \textbf{Methods} & \textbf{VarPointsTo$^*$}& \textbf{Throws$^*$}  \\ \hline 
   antlr &   \multirow{2}{*}{35,000} &  pdxfa+1obj  & 4.1x  &  1.3x  &  1.2x    &  1.5x &  2.8x
  \\
      & & pdxfa + eagc & 3.9x  & 1x & 1x &  3x&  4.6x    \\ 
     \hline 
         bloat &  \multirow{2}{*}{ 70,344}&  pdxfa+1obj  & 1.9x  & 1.4x  & 2.4x & 3.3x&  2.4x
          \\ &&  pdxfa + eagc  & 1.2x  &  1.3x & 1.1x &   6.3x & 6x      \\  % time: 10min
          \hline 
           chart &  \multirow{2}{*}{ 217,788}&  pdxfa+1obj  & 2.3x  & 1.3x & 1.1x & 2x &  2.3x
           \\ &&   pdxfa + eagc  & 2.1x  & 1.1x  & 1.2x &  6x  & 4.5x  \\  % & time   29 minutes.
            \hline 
            fop &  \multirow{2}{*}{ 184,386}&  pdxfa+1obj  & 2.1x &  1.4x  & 1.1x  & 4.2x &  5.5x
             \\& & pdxfa + eagc &  1.9x &  1.3x & 1.5x &  7.3x & 11x      \\ 
               \hline 
%              jython &  \multirow{2}{*}{ 110,867}&  pdxfa+1obj  & -  & -  & - &  - & -
%               \\ &&   pdxfa + eagc  & - &  - &  - & -& -   \\ 
%                  \hline 
                 hsqldb &  \multirow{2}{*}{ 155,591}&  pdxfa+1obj  & 8.9x  & 4.4x  & 3.4x &1x &  2.3x
                  \\ &&   pdxfa + eagc  & 5.3x  & 2.7x   & 3.3x & 3x &  4.5x  \\ 
                                                                                               \hline 
             luindex &  \multirow{2}{*}{ 38,221}&  pdxfa+1obj  & 1.9x  & 1.9x  & 1.8x & 1x & 1.6x 
                                        \\ &&   pdxfa + eagc  & 3.5x  & 1.7x  & 1.2x & 1.5x & 4x     \\ 
                                        \hline 
          lusearch & \multirow{2}{*}{87,000} &  pdxfa+1obj  & 1.5x  & 1.6x     & 1.6x   &   1.6x & 2.3x 
          \\
          & & pdxfa + eagc &  1x  & 1.5x  & 1.4x &  2.5x& 4.5x    \\ 
         \hline
          pmd & \multirow{2}{*}{55,000} &  pdxfa+1obj  & 1.8x  & 1.3x  & 1.5x & 2.2x & 5.2x
                  \\ && pdxfa + eagc  & 1.5x  & 1.1 x & 1x  &  3.7x & 7.7x \\ 
                  \hline
          xalan & \multirow{2}{*}{159,026} &  pdxfa+1obj  & 1.9x  & 1.3x & 1.7x & 2.8x &  6.2x
             \\ && pdxfa + eagc  & 1.4x  & 1.2x  & 1.3x &  3.7x & 10.3x  \\ 
                  \hline    
    \end{tabular}
    }
    \caption{\small
     Precision comparison. Values in columns \texttt{Nodes}, \texttt{Edges} and \texttt{Methods} are ratios of the 
number of nodes, edges and methods reached in our analysis, relative to the ones in WALA respectively. Values in 
columns \texttt{VarPointsTo$^*$} and \texttt{Throws$^*$} are ratios of   average cardinality of general point-to 
set and exception points-to set in WALA, relative to the ones in our analysis receptively. 
       Note that we did not list the results for the benchmark \texttt{jython} because it runs out of memory after one hour.
       %since we got OutofMemory error when running WALA after roughly one hour, even though we increased the stack and heap space in  JVM with the options:\texttt{Xms10g -Xss5g -Xmx10g -XX:MaxPermSize=2048m}. \texttt{pdxfa+1obj} exists to show the contribution of precision for \texttt{pdxfa} and \texttt{eagc} respectively.
     \textit{The table shows that the pushdown exception-flow analysis with enhanced abstract garbage collection
\texttt{pdxfa+eagc}
    outperforms  finite-state  analysis in WALA
    in  precision by 4.5X-11X for \texttt{Throws} and up to 7X for general points-to information \texttt{VarPointsTo}.
 }
    }
     \label{tbl:precision-result}
     \vspace{-3mm}
\end{table*}

%% file: benchmarks-table-time.tex
%
%% % % current results, I will consider to add time. too.
%\begin{table}
%\centering
%{\footnotesize
%    \begin{tabular}{ c| c | c| c | c | c | c | c | c | c| c| }
%    \textbf{Benchmark}  & \textbf{opts} & \textbf{antlr} & \textbf{bloat} & \textbf{chart} & \textbf{fop} & \textbf{hsqldb} & \textbf{luindex} & \textbf{lusearch}   & \textbf{pmd}  & \textbf{xalan}  \\ \hline 
%     \multirow{2}{*}{} &  pdxfa+1obj  & 13.3x & 9.8x & 25.9x   &  12.5x & 16.3x  & 7.6x & 6.7x & 18.4x& 11.5x
%  \\
%       & pdxfa + eagc &   11.7x  &  5.9x &14.9x & 10.9x & 5.4x  & 2.7& 5.7x & 9.2x &  5.0x \\ 
%     \hline 
%    \end{tabular}    
%    }
%    \caption{Benchmark results: {\tt{VarPointsTo}} and {\tt{Throws}} is 
%    presented as tuples $(a,b)$, where $a$ is the total entries, $b$ is the average 
%    types being invoked on in {\tt{VarPointsTo}} case,
%     and average exception objects thrown in {\tt{Throws}} case. 
%     All times are in seconds.
%     $\infty$ denotes the analysis did not finish within 6000 seconds.}
%     \label{tbl:precision-result}
%\end{table} 

\begin{table*}
\centering
{\footnotesize
    \begin{tabular}{ c|c|c|c|c|c|c|c|c|c}
    \hline
    \textbf{Benchmark} & \textbf{antlr}
                     & 
    \textbf{bloat}   & 
    \textbf{chart}   & 
    \textbf{fop}     & 
    \textbf{hsqldb}  & 
    \textbf{luindex} & 
    \textbf{lusearch}& 
    \textbf{pmd}     & 
    \textbf{xalan}   \\
    \hline
   \textbf{Ratio} & 8.5x & 5.6x & 9.7x & 7.9x & 5.2x & 3.1x & 8.7x & 9x & 4.7x \\
    \hline
    
    \end{tabular}    
    }
    \caption{Analysis time}
     \label{tbl:result-time}
\end{table*} 